\documentclass[aps,prd,twocolumn,10pt,floatfix,altaffilletter,preprintnumbers,showpacs,showkeys,nofootinbib,notoccite]{revtex4-2}

\usepackage{amsmath}
\usepackage{graphicx}
\graphicspath{{figs/}}
\usepackage{dcolumn}
\usepackage{bm}
\usepackage{xcolor}
\usepackage{tikz}
\usepackage{subfigure}
\usepackage{amsfonts} 
\usepackage{hyperref} 
\usepackage[T1]{fontenc}
\usepackage{url}
\usepackage{booktabs}
\hypersetup{
  colorlinks,
  citecolor=blue,
  linkcolor=purple,
  urlcolor=blue
  }

\newcommand{\mA}{{\mathcal{A}}} 
\newcommand{\os}{{\omega_{{S}}}}    
\newcommand{\rchi}{{\chi^2/{\rm d.o.f.}}}

\definecolor{red}{rgb}{1.0,0.0,0.0}

\usepackage{tikz,xcolor,hyperref}
\definecolor{lime}{HTML}{A6CE39}
\DeclareRobustCommand{\orcidicon}{\hspace{-1mm}
	\begin{tikzpicture}
	\draw[lime, fill=lime] (0,0) 
	circle [radius=0.16] 
	node[white] {{\fontfamily{qag}\selectfont \tiny \,ID}};
	\draw[white, fill=white] (-0.0525,0.095) 
	circle [radius=0.007];
	\end{tikzpicture}
	\hspace{-3mm}
}
\foreach \x in {A, ..., Z}{\expandafter\xdef\csname orcid\x\endcsname{\noexpand\href{https://orcid.org/\csname orcidauthor\x\endcsname}
			{\noexpand\orcidicon}}
}

\begin{document}
\preprint{TIFR/TH/26-17}

\title{Parameterizing the Standing Accretion Shock Instability\\ for Inference with Galactic Supernova Neutrino Signals at IceCube}

\author{Dwaipayan Mukherjee\orcidB{}}
  \email{dwaipayan.mukherjee@tifr.res.in}

\author{Mohamed Rameez\orcidC{}}
 \email{mohamed.rameez@tifr.res.in}

\author{Basudeb Dasgupta\orcidA{}}
 \email{bdasgupta@theory.tifr.res.in}

\affiliation{Tata Institute of Fundamental Research, Homi Bhabha Road, Colaba, Mumbai 400005, India}
 
\date{June 29, 2026}

\begin{abstract}
Simulations of core-collapse supernovae have revealed an epoch of hydrodynamic instability in which the matter of the collapsing star undergoes quasi-periodic oscillations, known as the standing accretion shock instability (SASI). Neutrinos produced in the core of the star travel through this oscillating matter, and information about this epoch is encoded in their high-statistics event rate observable at neutrino observatories. We propose a parametrization of the SASI-modulation to study its broad features, enabling statistical inference of SASI parameters. For the benchmark Galactic supernovae considered, we show that {\sc IceCube} can identify this epoch of instability and reconstruct its parameters with precision at the sub-percent level for the SASI frequency, percent level for the peak time, and a few to ten percent level for the amplitude and duration.

\end{abstract}


\maketitle
\tableofcontents

\section{Introduction}
Core-collapse supernovae (CCSNe) are among the most intense sources of neutrinos in the Universe, emitting copious amounts of neutrinos and antineutrinos of all flavors over timescales of a few seconds~\cite{Janka:2017vlw,raffelt1996,Janka:2016fox,Janka:2006fh}. However, Galactic supernovae (SNe) are rare~\cite{Rozwadowska:2020nab}, with only one confirmed detection of SN neutrinos~\cite{Bionta_SN1987A} from SN1987A, which resulted in about two dozen events at a few detectors~\cite{Hirata:1988ad,Kamiokande-II:1987idp,ALEXEYEV1988209}. Despite the low statistics, numerous studies~\cite{Burrows:1987zz,Arnett1989SN1987A_anuual_rev,Herant_1991,Vissani:2008ac,Utrobin_2015,Fiorillo:2023frv, Li:2023ulf} of the neutrino signal of SN1987A have validated the general picture of stellar core collapse and placed constraints on a variety of theoretical ideas. These neutrinos travel almost undeflected from their original path and carry the information about the internal dynamics of the dying star. Currently operating neutrino detectors like {\sc IceCube}~\cite{IceCube:2011cwc,IceCube:2016zyt,IceCube:2023ogt} and KM3NeT~\cite{KM3NeT:2009xxi,KM3NeT:2021moe}, and next-generation detectors such as {\sc IceCube}-Gen2~\cite{IceCube-Gen2:2020qha}, HyperKamiokande~\cite{HyperKamiokande, Hyper-Kamiokande:2021frf}, and DUNE~\cite{DUNE:2020lwj,DUNE:2024ptd}, will detect neutrinos from a Galactic SN with high statistics~\cite{Adams:2013ana,Li:2020ujl}. These detectors employ different interaction channels, which can offer complementary information about the neutrino energy, flavor conversions, and nuclear processes occurring in the PNS environment. A comprehensive review of neutrinos from CCSNe can be found in Refs.~\cite{Mirizzi:2015eza,Raffelt:2025wty}.

State-of-the-art simulations of CCSNe \cite{Fryxell_2000,Rampp:2002bq,Liebendoerfer:2002xn, Buras:2005rp,Fischer_2010,Skinner:2018iti,Bruenn:2018wpz,OConnor:2018sti,Nagakura:2020qhb,Akaho:2020xgb} show that hydrodynamic instabilities arise during the accretion phase when the shock is stalled at a radius of $R_{\rm sh}\!\sim150-200\,{\rm km}$, above the proto-neutron star (PNS) surface of radius $R_{\rm PNS}$. These large-scale instabilities, termed standing accretion shock instability, have been observed by several groups~\cite{Foglizzo:2002hi,Blondin:2002sm,Ohnishi:2005cv,Blondin:2006yw,Blondin_2007,Scheck:2007gw,Iwakami:2007ie,Iwakami:2008qj,
Fernandez:2010db,Hanke:2011jf,Foglizzo:2006fu,Foglizzo_2012,Tamborra:2013laa,Hanke:2013ena,Hanke_PhD_Thesis,Muller:2014rpb,Tamborra:2014aua,Tamborra:2014hga,Vartanyan:2019ssu,Choi:2025igp}, and are believed to be one of the ingredients for shock revival and successful explosion~\cite{Hanke:2011jf}. These sustained oscillations modulate the neutrino luminosities, imprinting quasi-periodic oscillatory features in the neutrino signal. These features in neutrino light curves encode crucial information on the location of the shock front and the dynamics of these instabilities~\cite{Muller:2014rpb}.

These oscillatory instabilities occur on timescales of 10–20\;ms~\cite{Muller:2014rpb,Muller:2020ard}, requiring detectors with sufficiently fine time-resolution. {\sc IceCube}'s supernova DAQ (SNDAQ) stores data in 2\;ms bins, while its planned upgrade, {\sc IceCube}-Gen2, is expected to be capable of sampling the SN signal down to 10\,ns resolution~\cite{IceCube-Gen2:2020qha}. {\sc IceCube} was designed to detect high-energy neutrinos over the GeV-PeV energy range; it can detect MeV-energy supernova neutrinos as a rise in ``noise'' across the detector, causing the entire detector to \emph{glow} due to the large flux~\cite{Halzen:1995ex,Kopke:2011xb}. Several works have studied the detectability of SASI-induced fast-time features in supernova neutrino light curves~\cite{Lund:2010kh,Beise:2023naa,Beise:2025njf}. Frequency-domain likelihood-ratio methods can identify SASI activity in {\sc IceCube} data and extract its frequency and amplitude~\cite{Lin:2019wwm}, while multi-messenger approaches combining neutrino and gravitational-wave signals have been studied to improve SASI identification in Galactic SNe~\mbox{\cite{Kuroda_2017,Lin:2022jea}}.

While previous studies have demonstrated that SASI-modulations are detectable, a simple functional description that can be directly incorporated into data-analysis pipelines is presently lacking. In this work, we propose a parameterized ansatz for the SASI-modulation of the rate of neutrino events at {\sc IceCube}. We validate our ansatz by fitting to the simulation data, treating each direction on the simulation grid as an independent source. We employ Monte Carlo sampling to extract SASI parameters for each line of sight. We discuss situations in which the fitting procedure does not converge and point to possible reasons. We use the SASI ansatz together with the {\sc IceCube} background noise to generate synthetic event rates and estimate the detection precision in extracting the SASI parameters. Finally, we discuss the broader implications of and the potential for detecting SN neutrinos with next-generation observatories.


\section{SASI Signal and Detection at {\sc IceCube}}
\label{Sec:SASI_mech_ansatz}


\subsection{SASI Mechanism}
\label{sec:Mechanism and Simulation}
Spherically symmetric simulations with detailed neutrino transport often fail to revive the stalled shock in massive stars~\cite{Mezzacappa:2000jb,Rampp:2000ws}. The widely accepted picture for shock revival is \textit{delayed neutrino-driven heating}, where neutrinos from the PNS deposit energy behind the stalled shock~\mbox{\cite{Bethe_SN_mechanism,Janka_CCSN_reflectionsdirection,Burrows:2012ew}}. This mechanism is aided by non-radial and multidimensional effects~\cite{Janka:2016fox, Nagakura:2020qhb}, one of which is known as SASI, first identified in 2D simulations~\cite{Blondin:2002sm}. SASI is a large-scale sloshing motion of the shock front that is shown to revive the shock~\cite{Hanke_PhD_Thesis, Ohnishi:2005cv}. SASI exhibits a clean periodicity of the shock surface and 
is typified by quasi-periodic oscillations sustained over a time scale of $\mathcal{O}(100\;{\rm ms})$. It is dominated by dipolar ($\ell=1$) and sometimes quadrupolar ($\ell=2$) modes, and has been shown to develop spiral modes in some 3D simulations~\cite{Blondin:2006yw,Blondin_2007,Fernandez:2010db,Hanke:2013ena}. SASI is interpreted as a vortical-acoustic cycle of the $\ell=1$ mode~\cite{Foglizzo:2002hi,Foglizzo:2006fu}. Here, small deformations of the shock front generate vorticity perturbations that are advected towards the PNS by the accretion flow. Near the PNS, where the flow decelerates, these perturbations produce outgoing sound waves, which propagate back and acoustically couple to the shock, forming a feedback cycle. Analogous instabilities have been observed in shallow water experiments~\cite{Foglizzo_2012}.

SASI is an oscillatory instability that has a typical periodic time scale $\tau_{\text{SASI}}$, which is a sum of the advective and acoustic crossing times, $\tau_{\rm adv}$ and $\tau_{\rm ac}$, between the shock and the deceleration region. Once the flow decelerates and becomes subsonic, the SASI time scale is dominated by the advection time scale. An expression for the $\ell=1$ mode time scale depends on the PNS radius and the shock radius as (with a weak dependence on the progenitor mass as $\sqrt{M}$)~\cite{Scheck:2007gw,Muller:2014rpb}
\begin{equation}
    \tau_{\text{SASI}}\approx\tau_{\rm adv}+\tau_{\rm ac}= \int_{R_{\rm PNS}}^{R_{\rm sh}}\frac{dr}{|v_r|} + \int_{R_{\rm PNS}}^{R_{\rm sh}}\frac{dr}{c_s - |v_r|},
\label{eq:sasi_time_scale}
\end{equation}
where $c_s$ and $v_r$ are the radius-dependent sound speed and
accretion velocity, respectively, in the postshock layer. Refs.~\cite{Tamborra:2013laa, Tamborra:2014hga} report the SASI frequency for both 20$\rm{M}_{\odot}$ and 27$\rm{M}_{\odot}$ simulations to be about 80\,Hz, as the $R_{\rm PNS}$ and $R_{\rm sh}$ are similar. In Ref.~\cite{Hanke:2013ena}, the shock initially exhibits a sloshing mode, but as the SASI amplitude grows, it develops spiral motion on a well-defined plane, which remains approximately stable. Simulations such as the \textsc{Flash} code by Fryxell et al. (2000)~\cite{Fryxell_2000}, the \textsc{Agile-Boltztran} code~\cite{Liebendoerfer:2002xn} and \textsc{Chimera} code~\cite{Bruenn:2018wpz} of the Oak Ridge-Basel group, \textsc{Fornax} code by Skinner et al. (2019)~\cite{Skinner:2018iti}, NSY Boltzmann code by the Waseda group~\cite{Akaho:2020xgb}, use different numerical approaches, but agree on the broad conclusions. A global comparison of several simulations is presented in Ref.~\cite{OConnor:2018sti}.
\begin{figure*}
    \centering
    \includegraphics[width=0.95\linewidth]{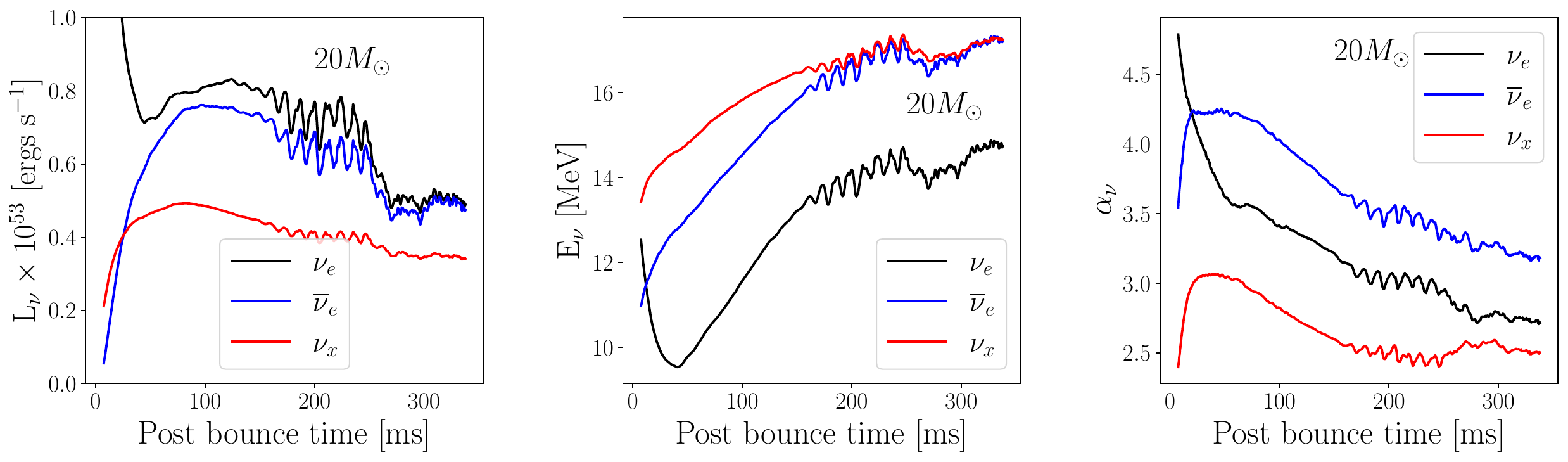}
    \vspace{1.9cm}
    \includegraphics[width=0.95\linewidth]
    {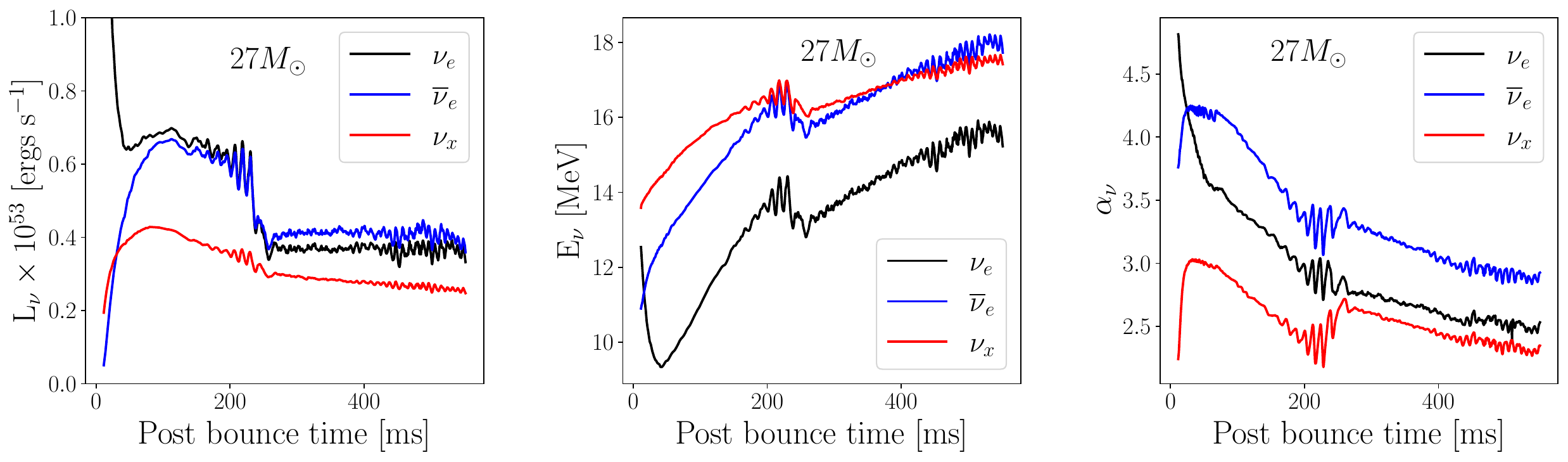}
   
    \caption{The panels show the luminosity (left), mean energy (middle), and pinching parameter $\alpha$ (right) for $\nu_e$, $\overline{\nu}_e$, and $\nu_x$ energy spectra as functions of post-bounce time, for the 20$\rm{M}_{\odot}$ (top) and 27$\rm{M}_{\odot}$ (bottom) stars along the \emph{violet} direction of Ref.~\cite{Tamborra:2014hga} (their Fig.\,6). The $\nu_e$ luminosity peaks at $\sim 3\times10^{53}\;{\rm erg\;s}^{-1}$ during the neutronization burst, while $\overline{\nu}_e$ and $\nu_x$ luminosities rise later during the accretion phase. The quasi-periodic SASI-modulations are visible in all three quantities in the time window $t_{\rm pb}\sim$150-250\;ms for both cases.
    }
    \label{fig:lum avgeng alpha for 20_27 violet}
\end{figure*}

For this paper, we focus on the 3D simulations of 20$\rm{M}_{\odot}$ and 27$\rm{M}_{\odot}$ stars~\cite{Hanke:2013ena,Tamborra:2014hga} (see also~\cite{Tamborra:2014aua}) by Tamborra, Raffelt, Hanke, Janka, and M\"uller~\cite{Tamborra:2013laa,Tamborra:2014hga}; the data is made available by the authors in the online archive of the Garching group~\cite{GARCHING_simulation, prvcommJanka}. Their neutrino-hydrodynamic simulation code \textsc{Prometheus-Vertex} combines a hydrodynamic solver and a neutrino transport module to solve for energy-dependent neutrino moment equations with up-to-date neutrino interaction rates~\cite{Rampp:2002bq, Buras:2005rp}. The models are computed on a spherical polar coordinate grid with a resolution of $n_r\times n_{\theta}\times n_{\varphi} = 400\times88\times176$. The neutrino flux seen by a distant observer is computed as a weighted hemispheric average over the emitting surface, where the contribution of each surface element is weighted by its projection along the observer's line of sight. The details of these flux-projection calculations are given in Sec.\;3 of Ref.~\cite{Muller:2011yi}.

The Garching simulation data files provide the projected luminosity $L_\nu(t)$, average energy $\langle E(t)\rangle$, and the second moment of energy $\langle E^2(t)\rangle$ as a function of post-bounce time, across the $n_{\theta}\times n_{\varphi} =88\times176$ directions on the simulation grid. Fig.\,\ref{fig:lum avgeng alpha for 20_27 violet} shows the luminosities, mean energies, and the pinching parameter for the 20$\rm{M}_{\odot}$ and 27$\rm{M}_{\odot}$ stars for a representative line of sight. We have verified this figure to be identical to Fig.\;6 in Ref.~\cite{Tamborra:2014hga}. In the simulations, sustained SASI activity begins at around 150\;ms for both cases, and ends after 250\;ms. For 27$\rm{M}_{\odot}$, there is a second SASI episode around 400\;ms, attributed to convective overturn, but in this work we focus only on the first SASI epoch.  


\subsection{Event Rate at {IceCube}}
\label{sec:Event rate in Icecube}
The {\sc IceCube} neutrino observatory is a ${\rm km}^3$-sized detector at the geographic South Pole, designed primarily to detect neutrinos with energies $\mathcal{O}(100\,{\rm GeV})$ and above. It consists of the main in-ice array, DeepCore, and IceTop, with 5160 digital optical modules (DOMs) deployed on 86 strings, each containing an array of 60 DOMs, embedded in the South Polar ice.  The photomultiplier tubes (PMTs) inside the DOMs are all oriented downward to bias the detection efficiency towards upgoing events~\cite{IceCube:2016zyt}. The DOMs detect and digitize Cherenkov photons emitted by charged leptons produced in neutrino interactions. The supernova neutrino signal in {\sc IceCube} is qualitatively different from the standard high-energy event signal, since the resulting charged leptons do not produce reconstructed tracks and travel approximately $x_e \simeq 0.579\,{\rm cm}\,E_e /{\rm MeV}$, i.e.\ only a few cm, much smaller than the DOM separation of 100\,m~\cite{IceCube:2013dkx}.

Thus, neutrinos from a Galactic supernova at distances of ${\cal O}(10-40)\,{\rm kpc}$  collectively increase the ``noise'' rates of individual DOMs over time scales of ${\cal O}(0.1-10)\,{\rm s}$. While the excess in a single DOM cannot be distinguished from noise, the correlated rise across the full detector becomes detectable. To detect this signal, {\sc IceCube} uses a dedicated data stream, the Supernova Scalers, in addition to the standard waveform-based acquisition, in which all discriminated PMT pulses are counted in intervals of $2^{16}/(40\,\mathrm{MHz}) = 1.6384\,\mathrm{ms}$ and analyzed online by the supernova data acquisition system (SNDAQ). The scaler information from individual DOMs is then synchronized and regrouped into 2\;ms bins for storage and processing.\footnote{Related methodological developments like the HitSpooling system in {\sc IceCube}, for improved studies of detector noise and atmospheric-muon backgrounds are studied in Chapters.\;4 \&\;5 of Ref.~\cite{HeeremanvonZuydtwyck:PhD_Thesis}, and the detector-response framework developed in Chapter. 4 of Ref.~\cite{BenediktRiedel:PhD_Thesis}, models the {\sc IceCube} response to Galactic supernova neutrinos including atmospheric muon backgrounds.} This makes {\sc IceCube} a unique detector as it provides information with a 2\;ms time resolution and $\mathcal{O}(10^2-10^3)$ Cherenkov photons per time bin.

At MeV energies, the main interaction channels for neutrinos are inverse beta decay (IBD), being the dominant channel with $\sim$ 93--94\% contribution to the signal~\cite{Strumia:2003zx}, $\nu-e^-$ scattering~\cite{Marciano:2003eq}, and $\nu-O$ interactions~\cite{Haxton:1987}, contributing 3--6\% of the signal. In our analysis, we compute the contribution from the IBD channel and rescale the rate by a factor of $1/0.93$ to account for other channels. The produced $e^{\pm}$ lose energy as they propagate through the ice, while emitting Cherenkov photons. The number of Cherenkov photons $N_{\gamma}$ is calculated by evaluating the Frank-Tamm equation, which gives the number of Cherenkov photons emitted per distance traveled, per wavelength, $d^2N_{\gamma}/dx\;d\lambda$. For a wavelength range of 300-600\,nm and a refractive index of $n = 1.32$, we get $N_{\gamma}= 178\,E_e$/MeV for positron energies $E_e<100\,{\rm MeV}$.

The large spacing between the strings and DOMs makes it hard to extract additional information, such as pointing information, spectral features, etc. Thus, the detection at {\sc IceCube} can only be seen as a counting experiment of the time-dependent rate given by (at a single DOM)~\cite{IceCube:2011cwc},
\begin{equation}
    \begin{aligned}
        R_{SN}(t) &= n_{\text{ice}} \int_0^{\infty} dE_{\nu} \Phi_{\nu}^{D} (E_{\nu},t ) \\
        &\times\int_{0}^{\infty} dE_e \dfrac{d\sigma}{dE_e} (E_{\nu},E_e) V_{\text{eff}}^{e^{\pm}}(E_e),
    \end{aligned}
    \label{eq:rate_SN_eqn}
\end{equation}
where $n_{\text{ice}} = 6.18\times 10^{22}\text{cm}^{-3}$ is the target density of protons in ice, and $E_e$ is the final-state positron energy. $d\sigma/dE_e$ is the IBD differential cross-section~\cite{Strumia:2003zx}. The effective detection volume for leptons $V_{\text{eff}}^{e^{\pm}}(E_e) =  N_{\gamma}(E_e) V_{\text{eff}}^{\gamma}(E_e)$, with the average effective volume for single-photon detection,  $V_{\text{eff}}^{\gamma} = 0.163\times 10^6\;{\rm cm}^3$~\cite{IceCube:2011cwc}. The total signal across all the DOMs is given by
\begin{equation}
    R(t) = N_{\rm DOM}~ \epsilon~ R_{SN}(t),
\label{eq:Total_rate_eqn}
\end{equation}
where $N_{\rm DOM}=5160$ is the total number of {\sc IceCube} DOMs, $\epsilon$ is the detection efficiency, and $R_{SN}(t)$ is the single-DOM supernova rate from Eq.\,\ref{eq:rate_SN_eqn}. The efficiency accounts for a non-paralyzing dead time $\tau_{\rm dt}=250\,\mu$s applied after every hit, which suppresses the effective rate by $\sim$13\% at a background rate of 286\;Hz per DOM: 
\begin{equation}
    \epsilon = \dfrac{0.87}{1+ R_{SN} \tau_{\rm dt}}. 
    \label{eq:epsilon}
\end{equation}

The energy- and time-dependent neutrino number flux at the detector is given by
\begin{equation}
    \Phi_{\nu}^{D}(E_\nu,t)
    =
    \frac{1}{4\pi d^2}
    \frac{L_{\nu}(t)}{\langle E_{\nu}\rangle(t)}
    f\big(E_{\nu},t\big),
    \label{eq:nu_flux_detector}
\end{equation}
where $d=10$ kpc is the fiducial distance to the supernova, $L_{\nu}(t)$ is the luminosity of a given neutrino flavor, $E_{\nu}(t)$ is the energy, and $\langle E_{\nu}\rangle(t)$ is the mean energy, all evolving with time.  

The energy spectrum $ f (E_{\nu})$ is taken to be~\cite{Keil:2002in},
\begin{equation}
   \begin{aligned}
        f\big(E_{\nu}\big) &= \dfrac{(\alpha +1)^{\alpha +1}}{\langle E\rangle \, \Gamma(\alpha +1)} \Bigg( \dfrac{E_{\nu}}{\langle E\rangle} \Bigg)^{\alpha}\times{\rm e} ^{-(\alpha+1)\dfrac{E_{\nu}}{\langle E\rangle}}\,,
   \end{aligned}
    \label{eq:Pinched_spectrum}
\end{equation}
introducing a pinching parameter $\alpha(t)$ that controls the high-energy tail of the spectrum.
The pinching parameter is calculated from the first and second moments of energy as 
\begin{equation}
    \alpha(t) = \dfrac{\langle E^2(t)\rangle - 2\langle E(t)\rangle^2}{\langle E(t)\rangle^2 -\langle E^2(t)\rangle}.
    \label{eq:pinching_parameter}
\end{equation}
Simulations suggest $2\leq \alpha(t) \leq 5$, depending on the flavor of the neutrino, as shown in the right panels of Fig.\,\ref{fig:lum avgeng alpha for 20_27 violet}.
Flavor conversion can modify the flavor fluxes before the neutrinos reach the detector. In general, the detected $\overline{\nu}_e$ flux can be written as
\begin{equation}
    \Phi_{\overline{\nu}_e}^{D}    =
    \overline{P}_{ee}\,
    \Phi_{\overline{\nu}_e}^{0}+
    \left[1-\overline{P}_{ee}\right]\,
    \Phi_{\nu_x}^{0},
    \label{eq:osc_flux}
\end{equation}
where $\Phi_{\overline{\nu}_e}^{0}$ and $\Phi_{\nu_x}^{0}$ are the emitted fluxes and $\overline{P}_{ee}$ is the $\overline{\nu}_e$ survival probability. In the standard MSW scenario~\cite{Wolfenstein:1977ue,Mikheyev_Smirnov:1985zog}, the survival probability depends on the neutrino mass hierarchy, with $\overline{P}_{ee}\simeq\cos^2\theta_{12}\simeq0.7$ for NH and $\overline{P}_{ee}\simeq0$ for IH~\cite{Dighe:1999bi}. Fast flavor conversions can further modify the survival probability and can equilibrate the flavor fluxes~\cite{Capozzi:2018rzl}. These effects can be incorporated by replacing the emitted flux in Eq.\,\ref{eq:rate_SN_eqn} by the oscillation-modified flux in Eq.\,\ref{eq:osc_flux} before computing the event rate. We do not include flavor-conversion effects and instead use the ``no-oscillation'' $\overline{\nu}_e$ flux.


\subsection{Scattering and Absorption in Ice}
\label{sec:Scattering and Absorption Effects in Ice}

The optical properties of the glacial ice at the South Pole are governed by the absorption length $\lambda_{\rm abs}$ and the effective scattering length $\lambda_{\rm scatt}$. At a wavelength of $\sim$400\,nm, which corresponds to the peak emission of Cherenkov light, {\sc IceCube} measurements indicate $\lambda_{\rm abs} \approx
100\text{-}200\,\mathrm{m}$ and $\lambda_{\rm scatt} \approx
20\text{-}40\,\mathrm{m}$, both varying with depth. These values are taken from {\sc IceCube}'s South Pole Ice (SPICE) models~\cite{AARTSEN201373,IceCube:2013ntj}. The refractive index of ice relevant for photon propagation is $n\approx 1.35$ at 400\;nm~\cite{AARTSEN201373}. These optical parameters determine the travel time and survival probability of photons emitted by leptons in neutrino-induced interactions. Scattering increases the effective path length, delaying photon arrival time. For a propagation distance $L\sim10\,\mathrm{m}$ and an effective scattering length $\lambda_{\rm scatt}=40\,\mathrm{m}$, the time delay is
\begin{equation}
    \Delta t_{\rm scatt}\approx\frac{n\,L^2}{c\lambda_{\rm scatt}}\approx 11\,{\rm ns}.
    \label{eq:scatt_delay}
\end{equation}
For characteristic SASI frequencies $\os\sim75-80\,{\rm Hz}$, the corresponding period is $\tau_{\rm SASI}\approx10-12\,{\rm ms}$, so this delay is negligible and does not smear the SASI signal.
In KM3NeT, the effective scattering length in seawater is longer, $\lambda_{\rm scatt}\approx300\,{\rm m}$~\cite{KM3NeT:2009xxi}, so this effect is even smaller.

Unlike scattering, absorption can attenuate the signal strength and reduce the photon count. The photon survival
probability over a path length $L$ is roughly given by
\begin{equation}
    P_{\rm surv} \approx \exp\!\left(-\frac{L}{\lambda_{\rm abs}}\right).
    \label{eq:psurv}
\end{equation}
For $\lambda_{\rm abs} = 150\,{\rm m}$ and a propagation distance of \mbox{$L = 10\,{\rm m}$}, Eq.\,\ref{eq:psurv} gives $P_{\rm surv} \approx 0.94$ (conservative estimate). For {\sc IceCube}'s SN neutrino detection, the dominant contribution comes from interactions close to each DOM, and this effect is small. For KM3NeT, $\lambda_{\rm abs}=60\,{\rm m}$ and $L=10\,{\rm m}$ gives \mbox{$P_{\rm
surv}\approx0.85$}. In both cases, this attenuation uniformly rescales both the SASI rate and the SASI-modulation amplitude. The detected rate is therefore
\begin{equation}
    R_{\rm det}(t) = R(t)\cdot P_{\rm surv}\,.
    \label{eq:rdet}
\end{equation}
Absorption mainly rescales the total event rate and does not change the SASI amplitude. Therefore, neither scattering nor absorption introduces substantial damping or smearing of SASI-modulations, and these effects give only sub-leading corrections to our analysis.


\section{Parameterizing SASI}
\label{sec:SASI at Icecube}

\subsection{SASI Ansatz}
\label{sec:SASI Ansatz}
\begin{figure}
    \centering
    \includegraphics[width=0.98\linewidth]{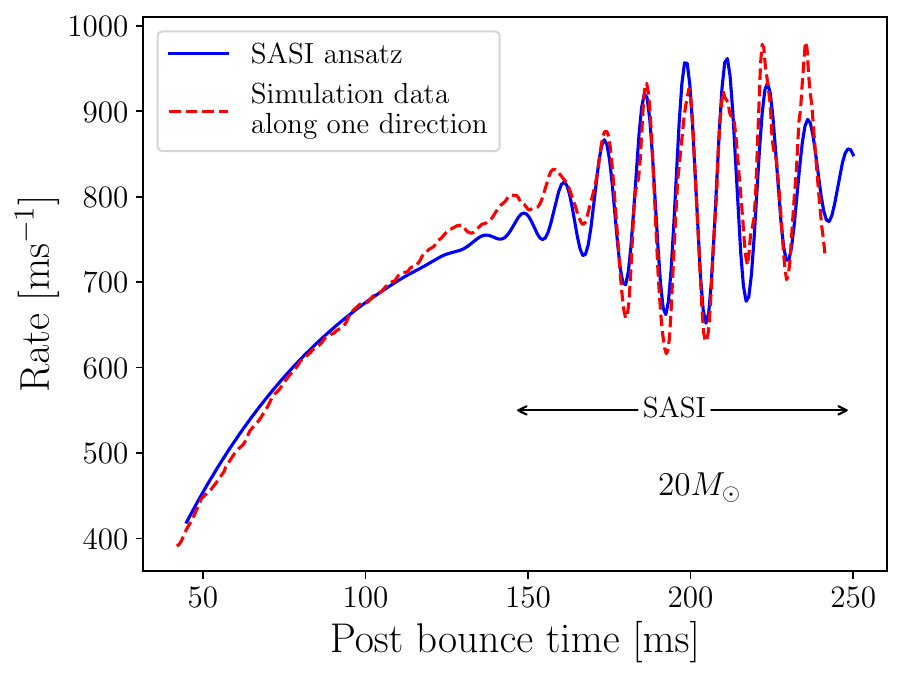}
    \includegraphics[width=0.98\linewidth]{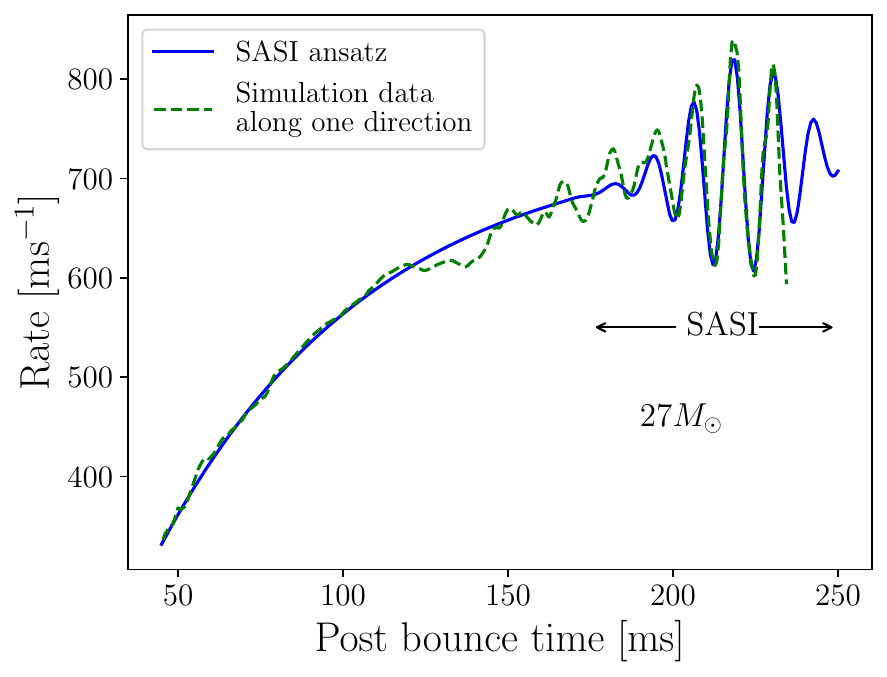}
    \caption{Reconstructed SASI waveforms from Eq.\,\ref{eq:fitting_function_eqn} (solid) compared with simulation data (dashed) along one representative line of sight, for 20$\rm{M}_{\odot}$ (top) and 27$\rm{M}_{\odot}$ (bottom) stars.}
    \label{fig: Reconstructed waveform}
\end{figure}

For a CCSN, the electron antineutrino ($\overline{\nu}_e$) luminosity begins to rise shortly after the neutronization burst, marking the onset of the accretion phase. This appears as an initial rise in the event rate until SASI-modulation ensues, as shown in the panels of Fig.\,\ref{fig: Reconstructed waveform}. Halzen and Raffelt~\cite{Halzen:2009sm} parameterized the initial rise epoch of the rate by, 
 \begin{equation}
    R_{\nu}(t) = R_{0} \Bigg(1- \exp \bigg[-\dfrac{(t-t_R)}{\tau}\bigg] \Bigg).
    \label{eq:rise_func}
\end{equation}
Here, the onset time $t_R$ determines the post-bounce time at which the rate begins to increase. The saturation value $R_0$ controls the behavior of the rate after $\sim$100\;ms when SASI effects become prominent.

SASI is a complex phenomenon; we simplify it by focusing on the discernible features of the modulated event rate:
\begin{equation} 
R(t) = R_{\nu}(t)\Bigg[ 1+ \mA \exp\!\bigg(\!-\frac{(t-t_{m})^2}{2w^2} \bigg)\sin(\os t) \Bigg], \label{eq:fitting_function_eqn} 
\end{equation}
where $R_{\nu}(t)$ is the rise function defined above. Broadly speaking, we approximate the quasi-periodicity of SASI by a pure sinusoidal wave of frequency $\os$, whose amplitude is modulated by a Gaussian envelope, accounting for the gradual onset and termination of the neutrino signal at {\sc IceCube}. The prefactor $\mA$ represents the fractional SASI-modulation relative to the rise function $R_\nu(t)$. The mean of the Gaussian, $t_m$, determines the position of the maximum SASI amplitude, henceforth called max-time, while the width $w$ describes the duration over which SASI-modulation is active.

We consider each line of sight as an independent observation. This allows us to study the dependence of SASI parameters on the relative orientation of the line of sight with respect to the SASI spiral plane. We sample the azimuthal direction uniformly by selecting every alternate value of $n_{\varphi}$ from the original simulation grid, which contains $88\times176$ zones, giving us a total of $88\times88=7744$ lines of sight.
 
\begin{figure*}
    \centering

    \begin{minipage}{0.325\linewidth}
        \centering
        \includegraphics[width=\linewidth]{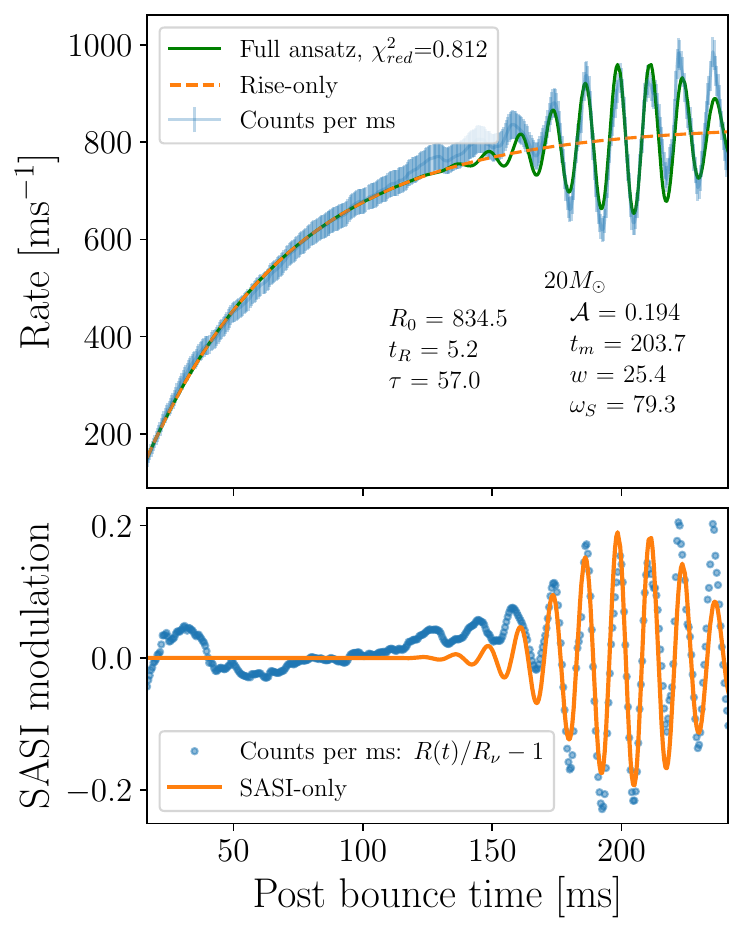}\\[-0.1ex](a)
    \end{minipage}
    \hfill
    \begin{minipage}{0.325\linewidth}
        \centering
        \includegraphics[width=\linewidth]{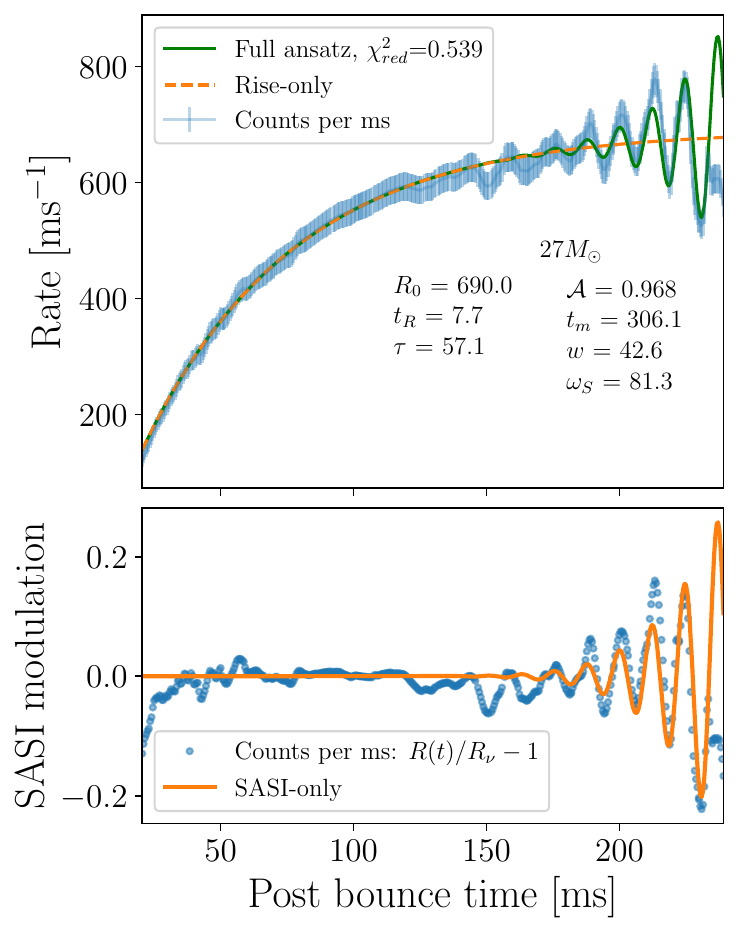}\\[-0.1ex](b)
    \end{minipage}
    \hfill
    \begin{minipage}{0.325\linewidth}
        \centering
        \includegraphics[width=\linewidth]{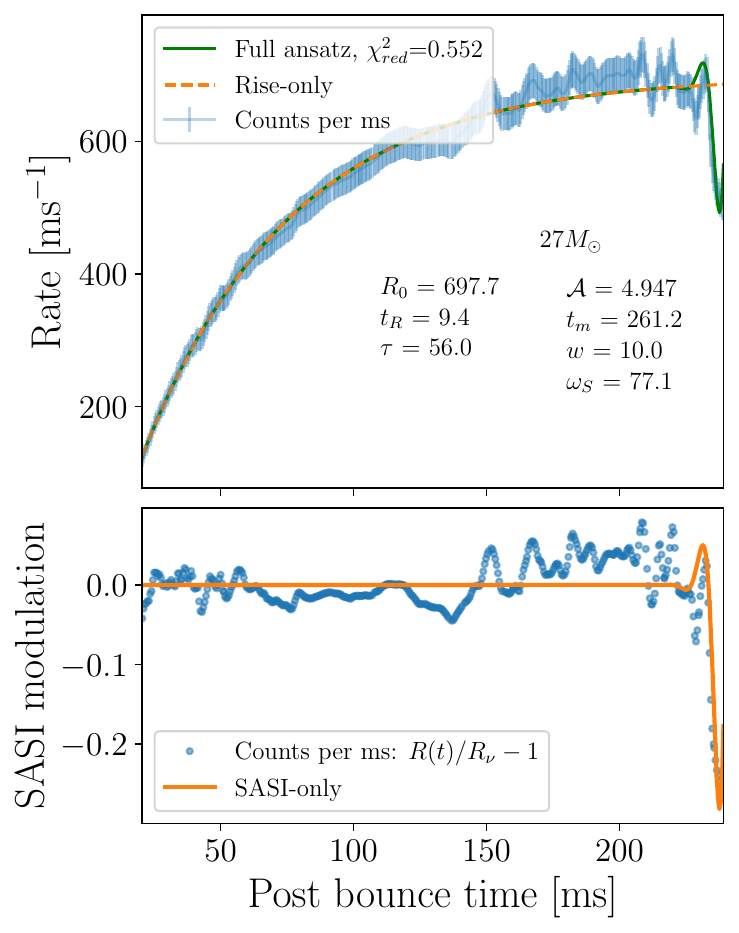}\\[-0.1ex](c)
    \end{minipage}

    \caption{Examples of good and bad fits. We see a good fit for 20$\rm{M}_{\odot}$ (a), and two poor fits for 27$\rm{M}_{\odot}$ (b) and (c). In each panel, the upper plot shows the full ansatz overlaid on the binned simulation data, shown as blue points with $\sqrt{N^{\rm bin}_{\rm count}}$ uncertainties, and the lower plot shows the isolated SASI-modulation $R(t)/R_\nu - 1$. Best-fit parameters are listed in each panel.}
    \label{fig:fitting_comparison}
\end{figure*}
We infer the parameters of Eq.\,\ref{eq:fitting_function_eqn} by fitting the calculated event rate for each line of sight for the 20${\rm M}_{\odot}$ and 27${\rm M}_{\odot}$ stars. To estimate the parameters, we assume uniform priors over physically motivated ranges for all parameters, which are $\mA\in[0,5]$, $w\in[1,120]$\;ms, $t_m\in[100,400]$\;ms, and $\os\in[50,100]$\;Hz. We adopt Poisson statistical uncertainties $\sigma_i=\sqrt{N^{\rm bin}_{\rm count}}$ in each time bin. In practice, we perform the estimation in two stages: first we determine the parameters of the rise function $R_{\nu}(t)$ by minimizing the least-squares statistic $\chi^2=\sum_i[(N^{\rm data}_i-N^{\rm model}_i)/\sigma_i]^2$ with the {\tt scipy} optimizer, and then, varying the rise parameters by 1$\sigma$ around their best-fit values, we sample the posterior distribution of the SASI-modulation parameters $(\mA, t_m, w, \os)$ using the Markov Chain Monte Carlo (MCMC) sampler implemented in the \texttt{emcee} package~\cite{MCMC_Foreman-Mackey_2013}. From these posterior samples, we take the posterior means as the best-fit parameters. To quantify the goodness-of-fit for each direction, we evaluate the reduced chi-square statistic, $\rchi$, for the full ansatz at the posterior mean, with ${\rm d.o.f.}=N-7$ for $N$ fitted time bins and seven total parameters; we consider the fit acceptable when $\rchi<1.5$ and the parameter values are within physically motivated limits, which are mentioned later.


\subsection{Ansatz Validation and Results}
In Fig.\,\ref{fig: Reconstructed waveform}, we validate our SASI-modulation ansatz (solid) against the simulation data (dashed) for both models, as seen by an observer along one line of sight. For 20$\rm{M}_{\odot}$ (top panel), the SASI ansatz in Eq.\,\ref{eq:fitting_function_eqn} describes the SASI features accurately. In particular, the amplitude and frequency are captured well near the peak of SASI activity, where the curves overlap well, and the oscillation frequency is nearly constant. For 27$\rm{M}_{\odot}$ (bottom panel), similar features are present, but the SASI activity is shorter-lived and less symmetric. The smaller amplitudes near the onset and the abrupt halt at the termination are poorly described by our ansatz.

Fig.\,\ref{fig:fitting_comparison} shows examples of fits. Panel~(a) shows a good fit for the 20$\rm{M}_{\odot}$ case, where the SASI amplitude and quasi-periodic variations are prominent. The inferred parameters lie within the expected range, and the fit gives $\rchi=0.812$, showing that the ansatz captures the main SASI features. Panels~(b) and (c) show two fit-failure cases for the 27$\rm{M}_{\odot}$ model. In panel~(b), the fit appears visually reasonable, but the inferred amplitude is close to unity, and the max-time is shifted to values larger than 250\;ms. This happens because the Gaussian envelope attempts to accommodate the abrupt termination of SASI activity, leading to artificially large values of $w$ and $t_m$. Panel~(c) shows a direction where the SASI-modulation is weak and lacks a quasi-periodic feature. Here, the fitted parameters are unphysical: for example, $\mA$ measures the SASI-modulation relative to the rise function $R_\nu(t)$, which is expected to remain below unity. Here, large amplitude, poorly inferred width, or shifted max-time therefore indicate unreliable fits. These examples show that a low $\rchi$ alone is not sufficient; the fitted parameters must also remain within physically reasonable ranges.

\begin{figure*}
    \centering
    \includegraphics[width=0.57\linewidth]{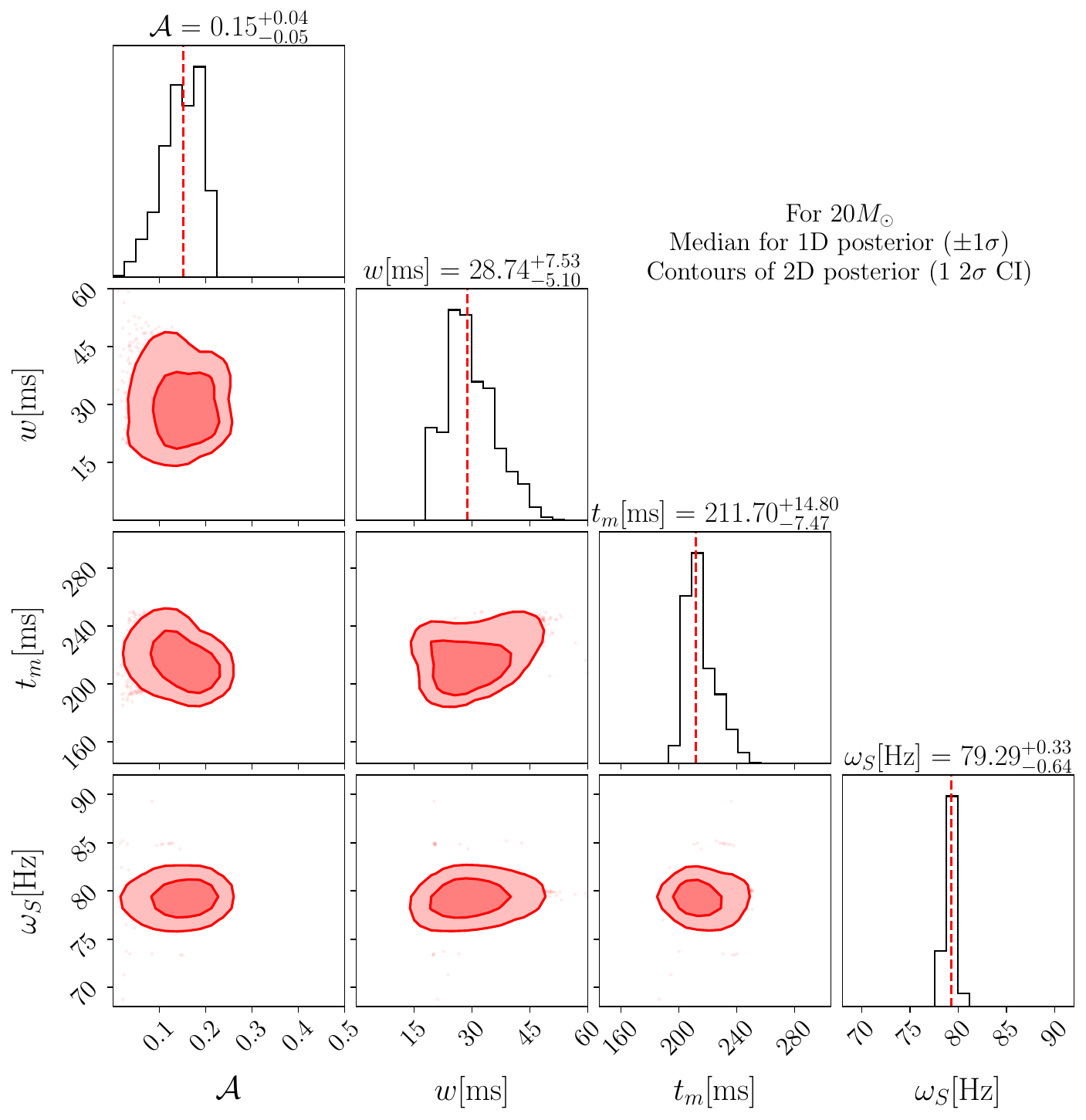}
    \includegraphics[width=0.57\linewidth]{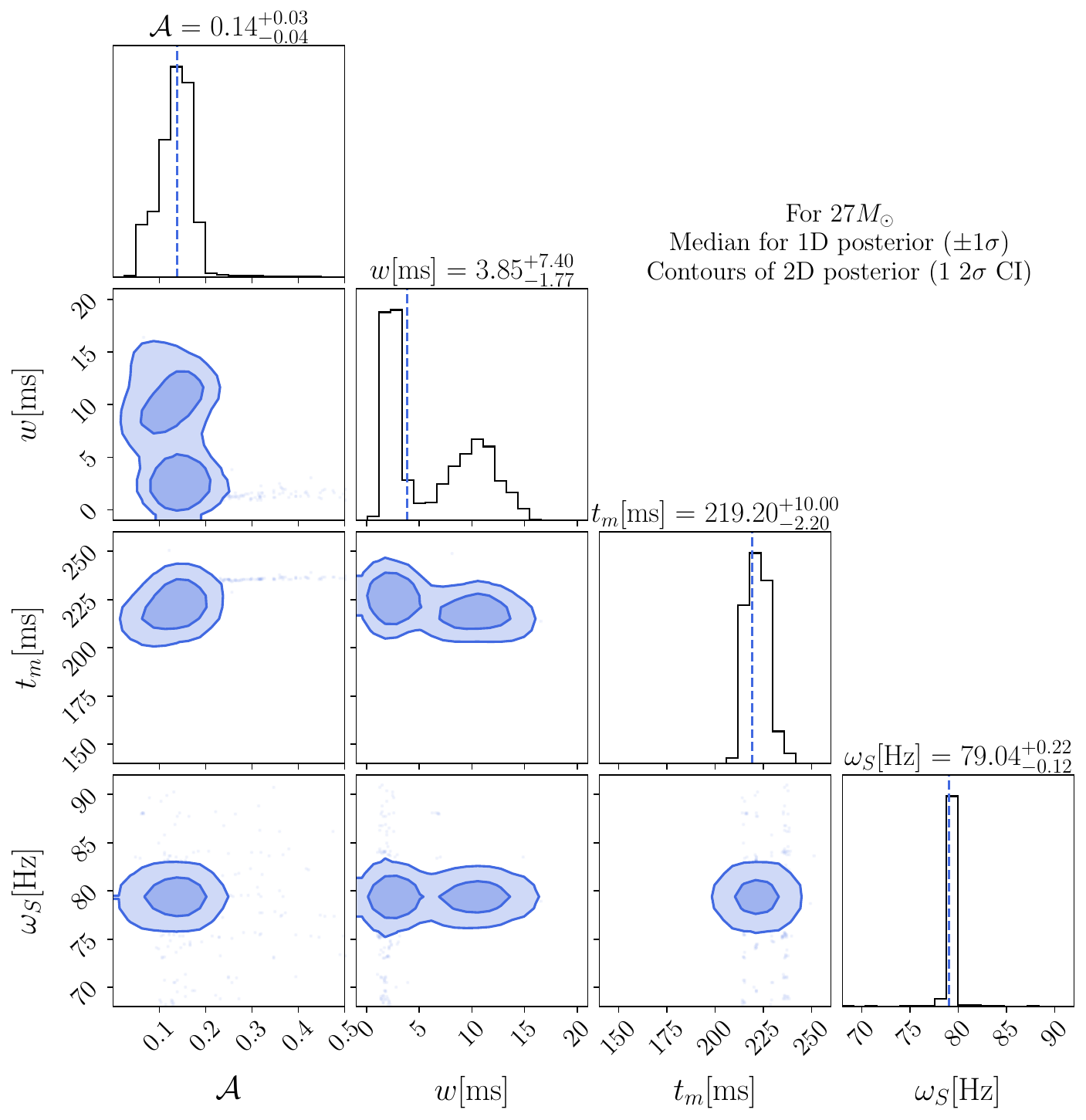}
    \caption{Corner plots showing the joint and marginalized posterior distributions of the SASI parameters ($\mA,\,w,\,t_m,\,\os$) across accepted directions, for the 20$\rm{M}_{\odot}$ case (top) with 7658 accepted directions and the 27$\rm{M}_{\odot}$ case (bottom) with 4533 accepted directions. The contours enclose the 1$\sigma$ and 2$\sigma$ regions, and the median values with 1$\sigma$ uncertainties are indicated. The frequency $\os$ is sharply peaked near 79\;Hz in both models. Other features of the distributions are discussed in the text.}
    \label{fig:SASI_parameters_histogram}
\end{figure*}

Fig.\,\ref{fig:SASI_parameters_histogram} shows the corner plots of the distributions of the four parameters of the SASI ansatz ($\mA, t_m, w, \os$). The one-dimensional (1D) posterior distributions, along with the two-dimensional (2D) joint distributions, are shown with 1$\sigma$ and 2$\sigma$ confidence intervals (C.I.) for both cases. The median values of the inferred parameters are derived from the posterior distributions and marked in the corresponding plots. From these panels, we draw the following key observations:
\begin{enumerate}
    \item The distributions of $t_m$ and $\os$ are sharply peaked for both models. The SASI activity reaches its maximum around $t_m\simeq 212$\;ms for 20$\rm{M}_{\odot}$ and $t_m\simeq 219$\;ms for 27$\rm{M}_{\odot}$. The frequency distribution is peaked near $\os\simeq79$\;Hz in both cases, consistent with previous findings~\cite{Tamborra:2014hga}. This weak dependence on viewing angle suggests that $t_m$ and $\os$ are controlled by the SASI dynamics and not by projection effects.

    \item  The distribution of amplitude $\mA$ is centered around 0.15 for 20$\rm{M}_{\odot}$ and 0.14 for 27$\rm{M}_{\odot}$, with larger values occurring less frequently. The spread in $\mA$ across viewing directions indicates that the inferred amplitude modulation depends on the line-of-sight projection with respect to the SASI spiral plane.
    
    \item The posterior distribution of $w$ shows different behavior in the two cases. For 20$\rm{M}_{\odot}$, the distribution is broad and peaked near 29\;ms, indicating a longer duration of SASI-modulation. For 27$\rm{M}_{\odot}$, the distribution has a sharp peak near 4\;ms, suggesting that SASI-modulation is short-lived. The inferred value of $w$ depends on how the onset and the abrupt termination of the oscillations are captured by the fitting procedure. If the onset modulation is weak relative to the peak amplitude and the termination is abrupt, then these regions may be averaged out, making the Gaussian envelope width harder to estimate.

    \item  We observe that for the 27$\rm{M}_{\odot}$ case, the width distribution has a second peak at larger $w$. This arises from directions where the modulation is weaker and not well described by a symmetric Gaussian envelope, causing the fit to shift $t_m$ and broaden the inferred width. This second peak is likely a fitting artifact. The directions where we see this constitute about 15\% of the accepted fits and appear as a small hump in the posterior distribution.
\end{enumerate}
\begin{figure*}[!htbp]
    \centering
    \includegraphics[width=0.49\linewidth]{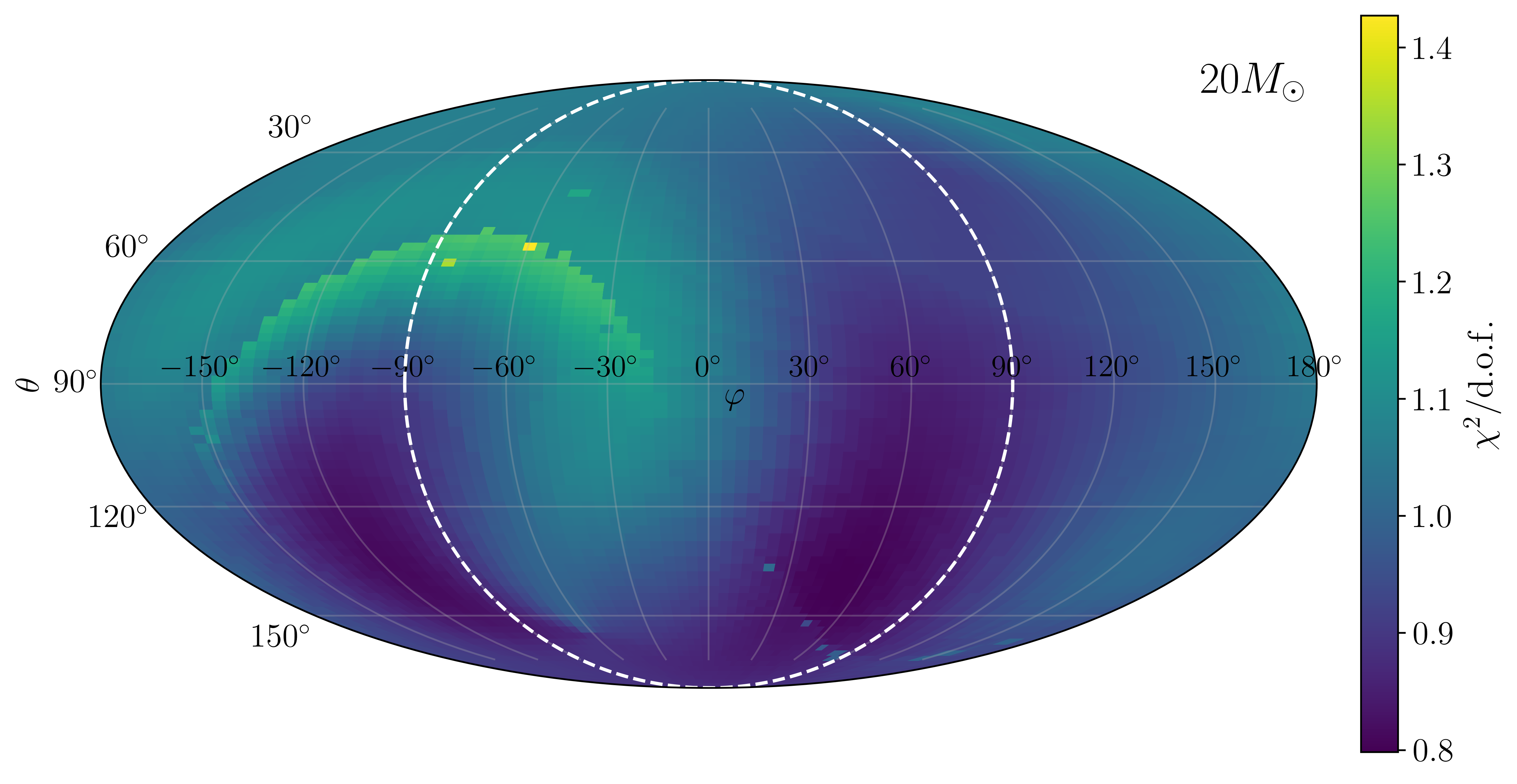}
    \includegraphics[width=0.49\linewidth]{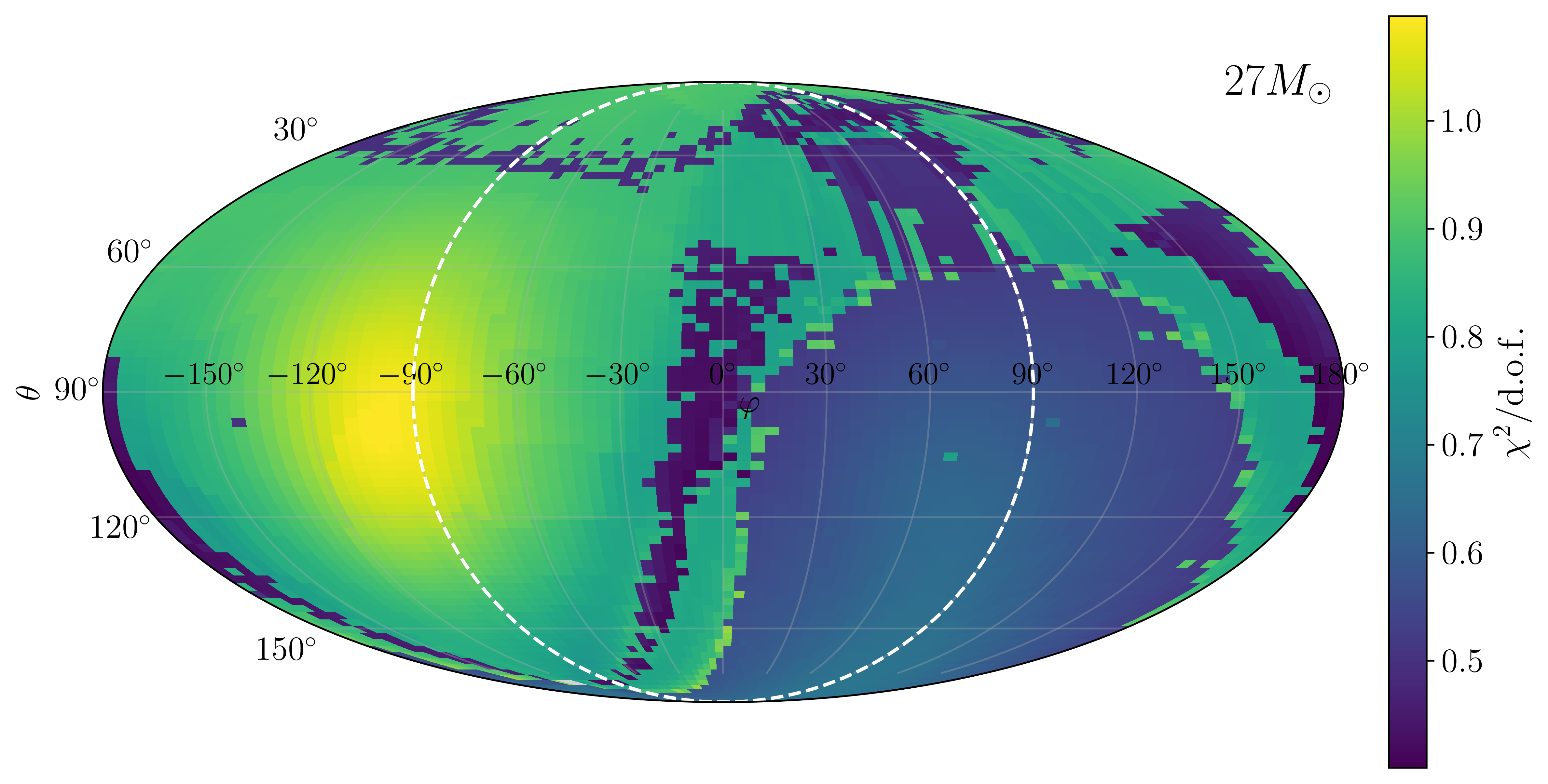}
    \caption{Sky maps of the reduced chi-square statistic $\rchi$ obtained by fitting the SASI ansatz (Eq.\,\ref{eq:fitting_function_eqn}) along different observer directions, for the 20$\rm{M}_{\odot}$ and the 27$\rm{M}_{\odot}$ stars. The dashed white great-circle indicates directions lying in the plane of the SASI spiral motion.
    }
    \label{fig:skymap_chisq}
\end{figure*}

\begin{figure*}[!htbp]
    \centering
    \includegraphics[width=0.49\linewidth]{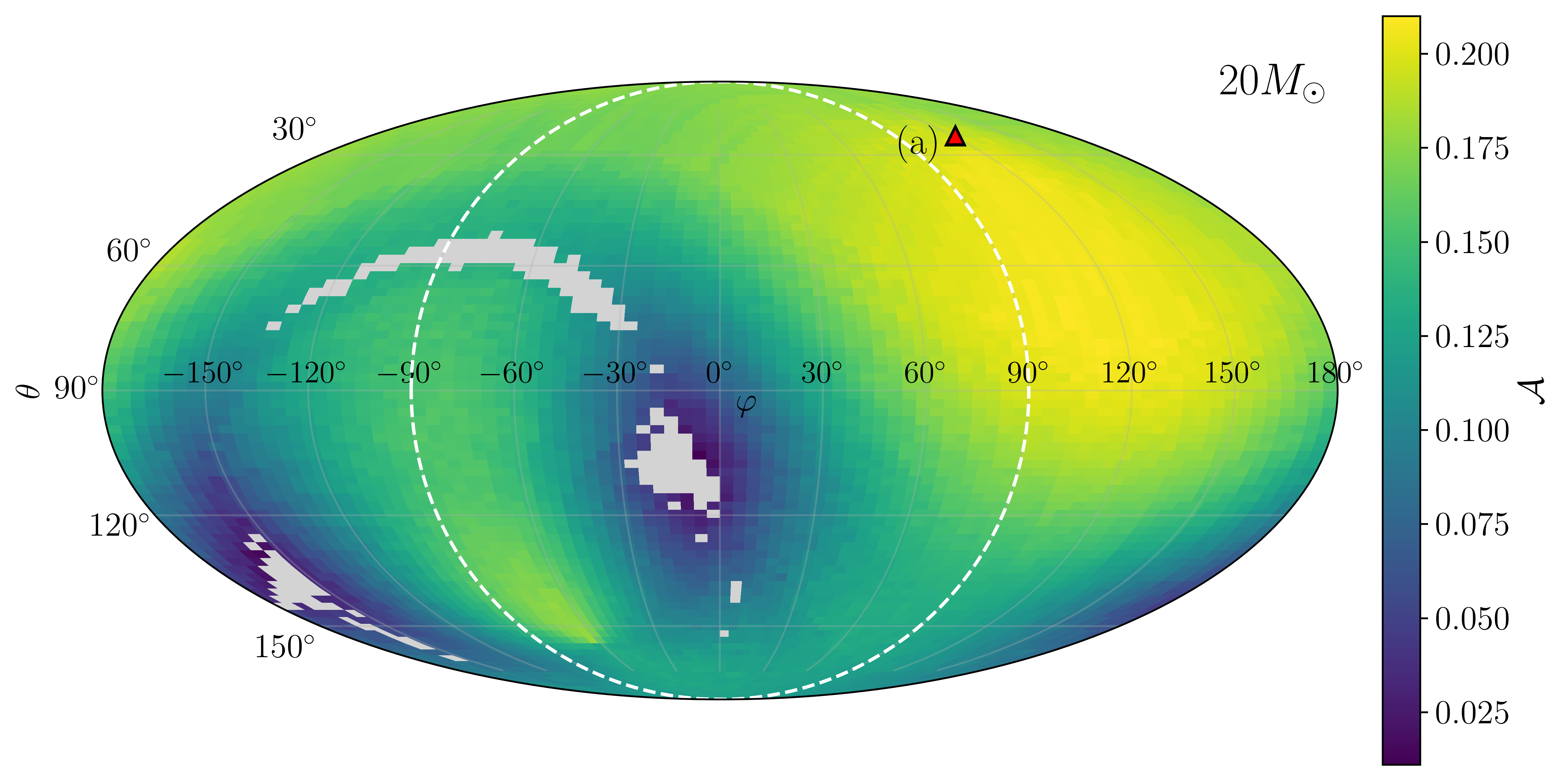}
    \includegraphics[width=0.49\linewidth]{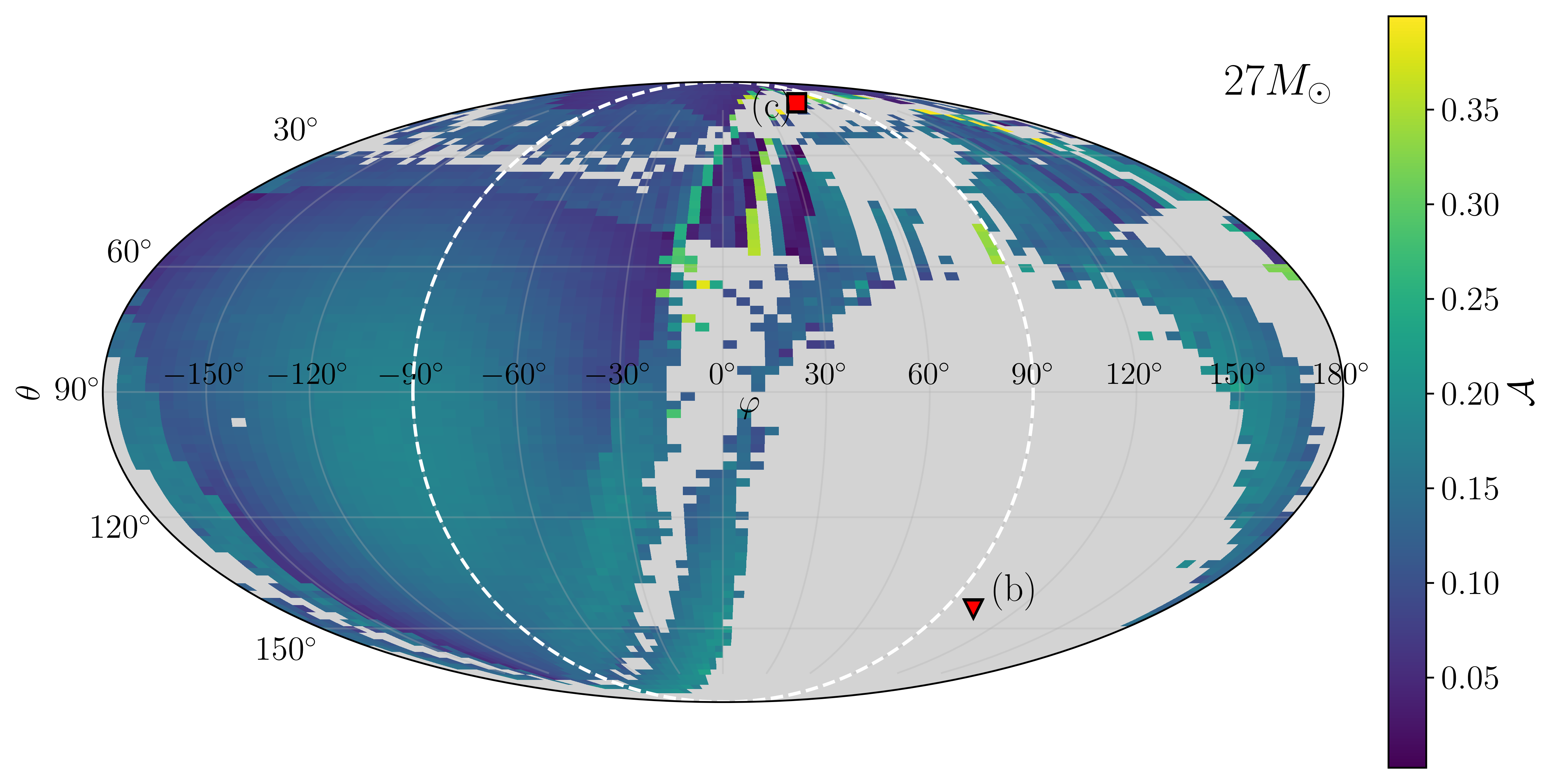}
    \caption{Sky maps of the fitted SASI amplitude $\mA$ for the 20$\rm{M}_{\odot}$ and the 27$\rm{M}_{\odot}$ stars, using the same sky orientation as in Fig.\,\ref{fig:skymap_chisq}. For 20$\rm{M}_{\odot}$, the amplitude is large for directions close to the SASI spiral plane, where the modulation is strongest. For 27$\rm{M}_{\odot}$, we see a more fragmented distribution. The grey regions indicate directions where the SASI parameters fall outside the parameter cuts. The markers in each panel represent the directions of the fitting examples shown in Fig.\;\ref{fig:fitting_comparison}. 
    }
    \label{fig:skymap_amp}
\end{figure*}

The parameter distributions shown in Fig.\,\ref{fig:SASI_parameters_histogram} are plotted using the following acceptance criteria: \mbox{({\it a})~$\mA<0.8$}, ({\it b})~$150<t_m<250$\;ms, ({\it c})~$w<80$\;ms, and \mbox{({\it d})~$60<\os<100$\;Hz}. The contours were constructed using only those directions where the fit parameters are within the above cuts and have $\rchi<1.5$ along them. For the 20$\rm{M}_{\odot}$ case, such fits were obtained for approximately 7658 directions, indicating that the functional ansatz given by Eq.\,\ref{eq:fitting_function_eqn} effectively describes the SASI phenomenon in the majority of cases. The high success rate is consistent with the behavior observed in the 20$\rm{M}_{\odot}$ case, where the SASI activity is prominent and sustained for a duration of 100\;ms. In contrast, for the 27$\rm{M}_{\odot}$ case, the fitting was successful in only about 4533 directions where the SASI parameters are within the above range. For poor-fit directions, the SASI activity is weaker and less sustained. This includes the directions with $\rchi\sim0.5$, where our ansatz overfits the weak SASI-modulation, resulting in SASI parameters outside the accepted range; e.g., see panel (c) of Fig.\,\ref{fig:fitting_comparison}.

In Refs.~\cite{Hanke:2013ena,Tamborra:2014hga}, the authors report that SASI develops a spiral mode associated with circular motion of the shock front within a plane, whose orientation is characterized by a unit vector $\hat{\bf{n}}$ normal to that plane. For the first SASI epoch of 27${\rm M}_{\odot}$, $\hat{\bf{n}}_{27} = (-0.35, 0.93, 0.11)$, and for the 20$\rm{M}_{\odot}$ star $\hat{\bf{n}}_{20} = (-0.56, -0.81, -0.20)$ in the SN simulation grid. We demonstrate the directional dependence of SASI relative to this plane, using Mollweide sky maps of $\rchi$ and the fitted SASI amplitude $\mA$ in Fig.\,\ref{fig:skymap_chisq} and Fig.\,\ref{fig:skymap_amp}, respectively. The sky maps are shown in the coordinate system of the simulation grid, rotated such that the origin denotes the direction of the unit vectors $\hat{\bf n}_{20,27}$ normal to the plane of spiral motion, while the dashed white great-circle denotes viewing directions lying in the plane.

For the 20$\rm{M}_{\odot}$ model, acceptable fits are obtained over most of the sky. The fitted amplitude $\mA$ shows larger values near regions close to the SASI spiral plane, while smaller values appear away from the plane, where the modulation is weak. The corresponding $\rchi$ values remain close to unity over a large fraction of the sky, indicating that the ansatz fits the simulation data well for most viewing directions. For the 27$\rm{M}_{\odot}$ model, the directional distribution is more fragmented. A large fraction of viewing directions is rejected by the parameter cuts, shown as grey regions in the sky maps. In the remaining accepted directions, $\mA$ varies significantly across the sky, with larger values appearing sparsely, rather than uniformly along the spiral plane. The $\rchi$ map similarly shows stronger directional variation.


\subsection{Light-Curves and Parameter Inference}
\label{sec:Extracting SASI params}
We model the event rate $R_{\rm obs}(\boldsymbol{\theta})$ at {\sc IceCube}, with parameters $\boldsymbol{\theta}=\{\mA,w,t_m,\os\}$, as
\begin{equation}
    R_{\rm obs}(\boldsymbol{\theta}) = \big(R(\hat{\boldsymbol{\theta}})+ R_{\rm bkg}\big) \{1+ \mathcal{N}(0,\sigma)\},
\end{equation}
where $R(\hat{\boldsymbol{\theta}})$ is the maximum-likelihood rate for the \emph{true} parameters $\hat{\boldsymbol{\theta}}$. ${\cal N}(0,\sigma)$ denotes the statistical fluctuation, and $R_{\rm bkg}$ is the total background rate across all DOMs. The inferred parameters deviate from their true values due to noise, and we quantify this effect.

\begin{table}[!h]
\centering
\renewcommand{\arraystretch}{1.2}
\setlength{\tabcolsep}{7pt}

\begin{tabular}{@{}c c c c@{}}
\toprule
$M\;[\mathrm{M}_\odot]$ & $R_0\;[\mathrm{ms}^{-1}]$ & $t_R\;[{\rm ms}]$ & $\tau\;[{\rm ms}]$ \\
\midrule
20 & 835 & 5 & 60 \\
27 & 711 & 9 & 57 \\
\bottomrule
\end{tabular}

\vspace{0.6em}

\begin{tabular}{@{}c c c c c@{}}
\toprule
$M\;[\mathrm{M}_\odot]$ & $\mA$ & $w\;[{\rm ms}]$ & $t_m\;[{\rm ms}]$ & $\os\;[{\rm Hz}]$ \\
\midrule
20 & 0.2 & 25 & 210 & 79 \\
27 & 0.2 & 15 & 219 & 79 \\
\bottomrule
\end{tabular}
\caption{The values of the SASI parameters used to generate Figs.~\ref{fig:Dist_of_extracted_amp_width} and~\ref{fig:Std_dev_SASI_params}. In the plots, when a particular parameter is varied, the others are held constant at these given values.}
\label{tab:rise_func_param_vals}

\end{table}

We bin the neutrino event rate at {\sc IceCube} into 1\;ms time intervals. In each 1\;ms time bin, if the number of detected events including background is $N_{\text{count}}^{\text{bin}}$, we sample the statistical noise from a normal distribution ${\cal N}(0,\sigma)$ with $\sigma=\sqrt{N_{\text{count}}^{\text{bin}}}$. We add an average {\sc IceCube} background noise of 286\;Hz at each DOM, giving a total background rate across all DOMs of $R_{\rm bkg}=1.48\times10^3\;{\rm ms}^{-1}$. Adding these fluctuations to all time bins gives a noise-added synthetic rate $R_{\text{obs}}(\boldsymbol{\theta})$, from which we infer the SASI parameters. To characterize the uncertainties, we repeat this procedure over $10^4$ Monte Carlo realizations of the observed {\sc IceCube} rate. We compute the synthetic event rates using the best-fit rise function and SASI parameter values from Fig.\,\ref{fig:SASI_parameters_histogram}, listed in Table\,\ref{tab:rise_func_param_vals}.

\begin{figure*}
    \centering
    \includegraphics[width=0.35\linewidth]{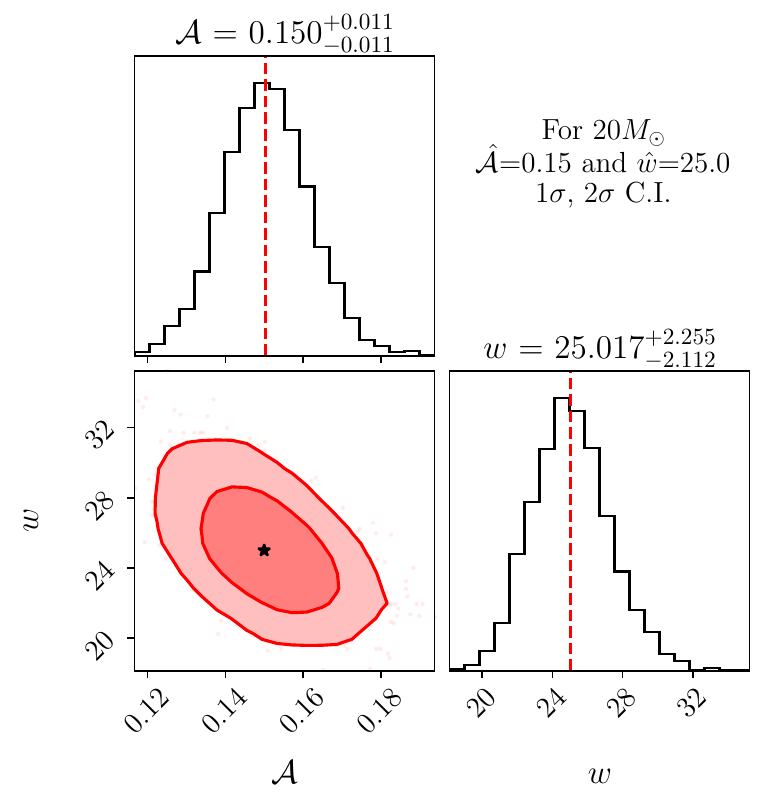}
    \includegraphics[width=0.35\linewidth]{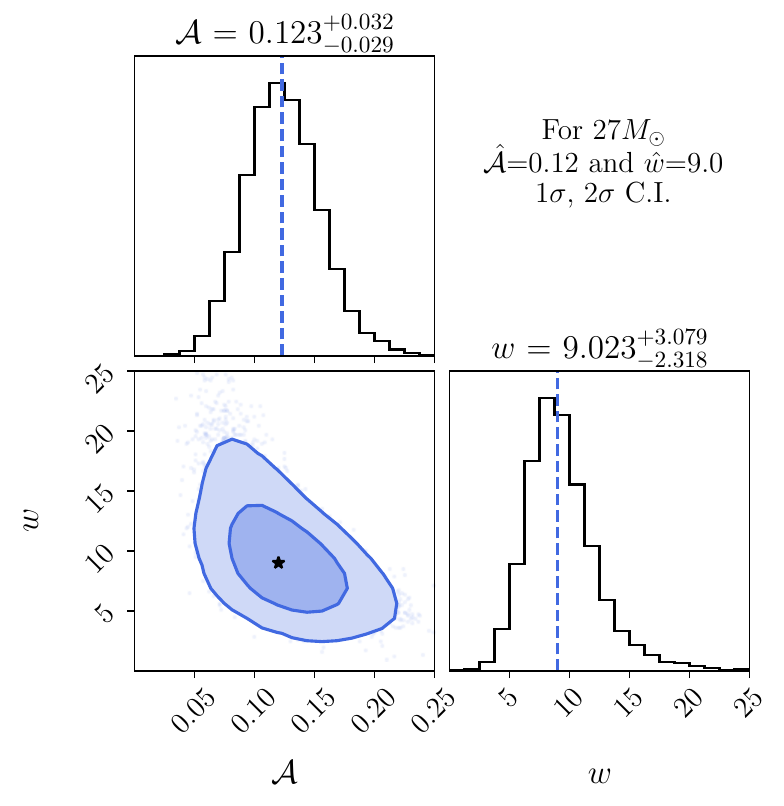}
    \caption{Joint posterior distributions of the extracted SASI amplitude $\mA$ and width $w$ from synthetic noise-added data of the {\sc IceCube} event rate, for 20$\rm{M}_{\odot}$ (left) with initial values $(\hat{\mA},\hat{w})=(0.15,\,25\,{\rm ms})$ and 27$\rm{M}_{\odot}$ (right) with $(\hat{\mA},\hat{w})=(0.12,\,9\,{\rm ms})$. Contours show $1\sigma$ and $2\sigma$ confidence regions, and the initial values are indicated by a star. $\os$ and $t_m$ have very narrow spreads in their distributions and are not shown here.}
    \label{fig:Dist_of_extracted_amp_width}
\end{figure*}

\begin{figure*}[!htbp]
    \centering
    \includegraphics[width=0.24\linewidth]{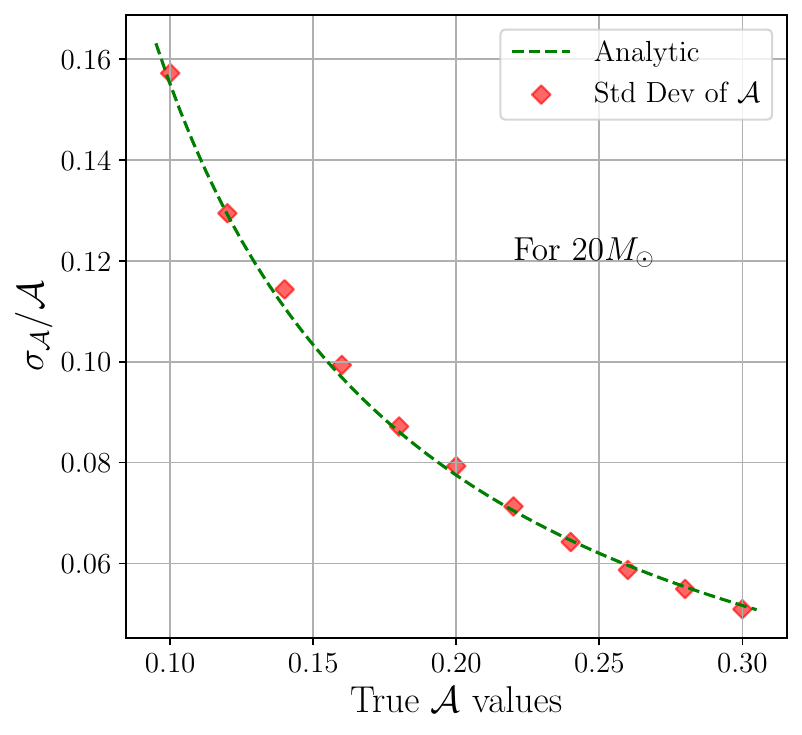}
    \includegraphics[width=0.24\linewidth]{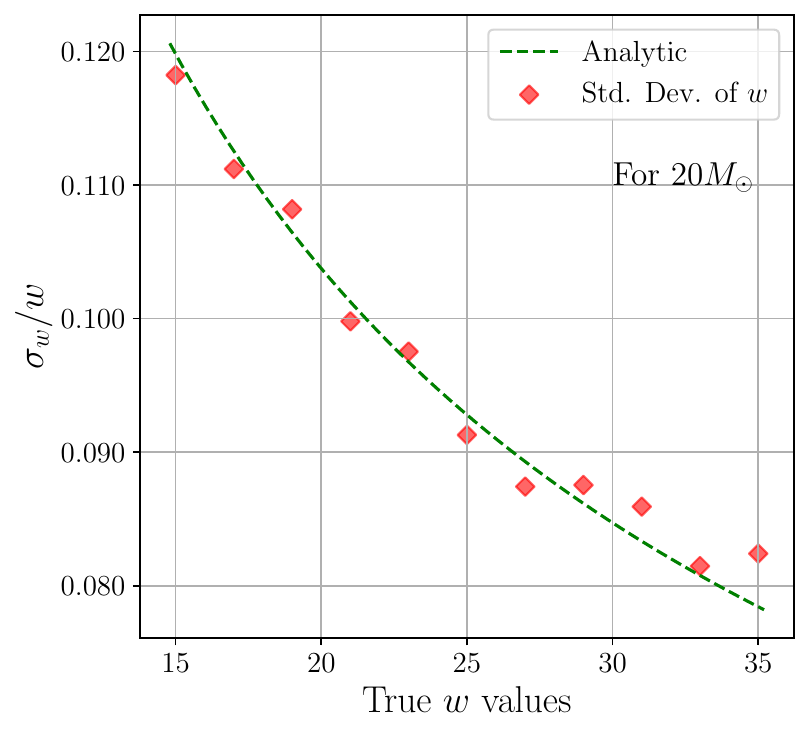}
    \includegraphics[width=0.24\linewidth]{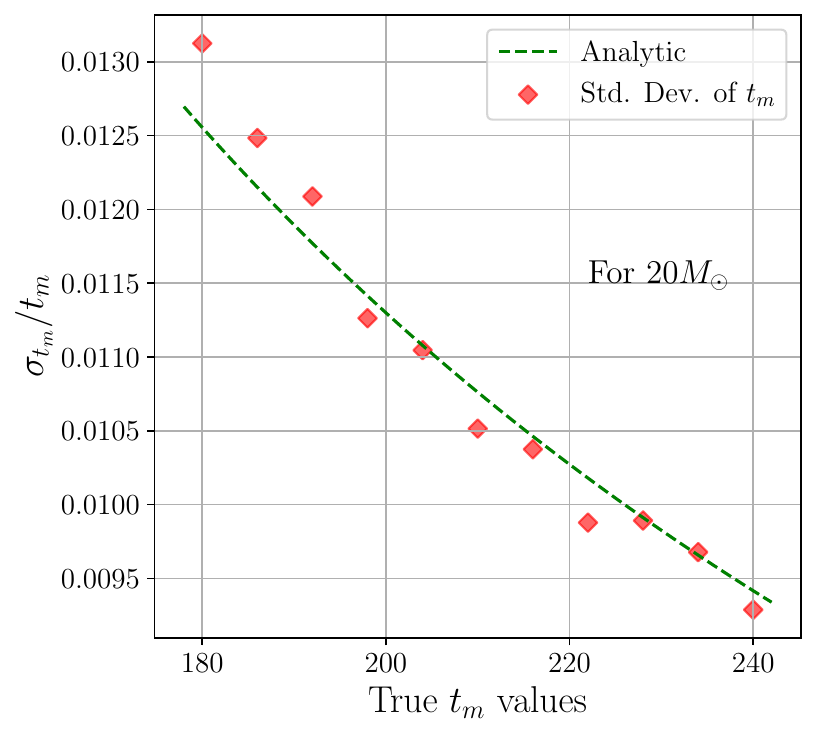}
    \includegraphics[width=0.24\linewidth]{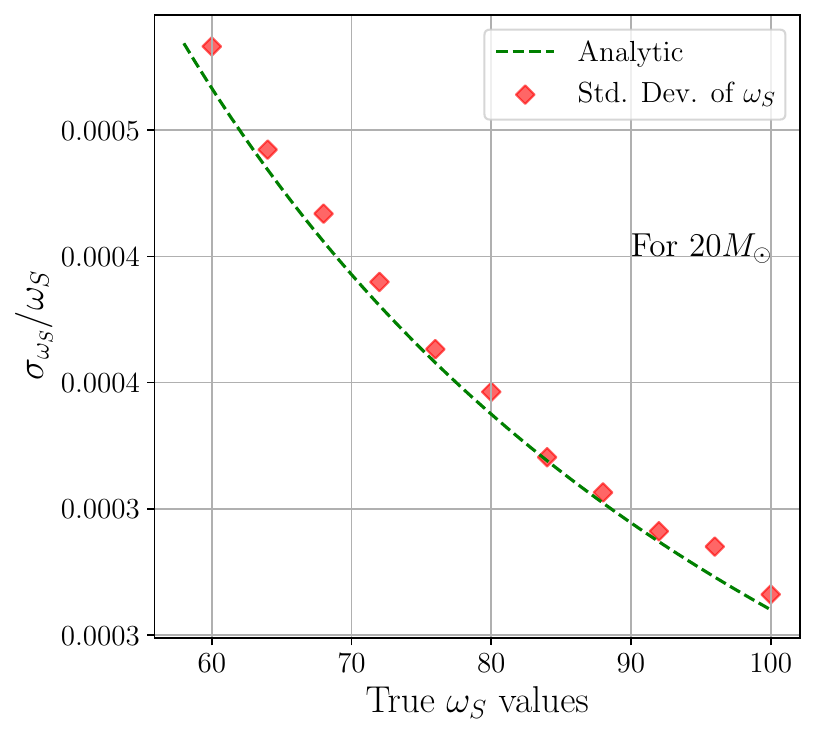}
    \includegraphics[width=0.24\linewidth]{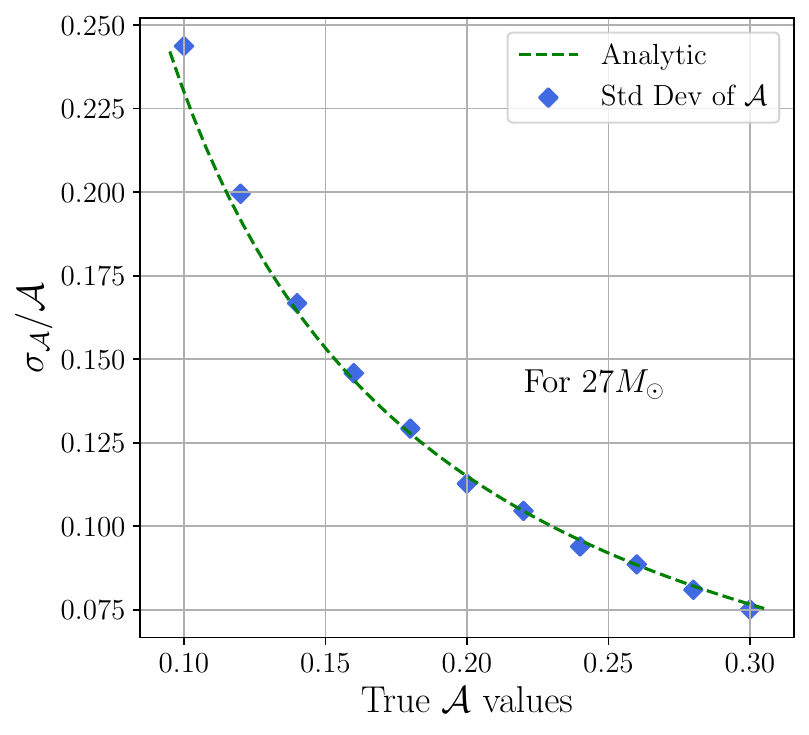}
    \includegraphics[width=0.24\linewidth]{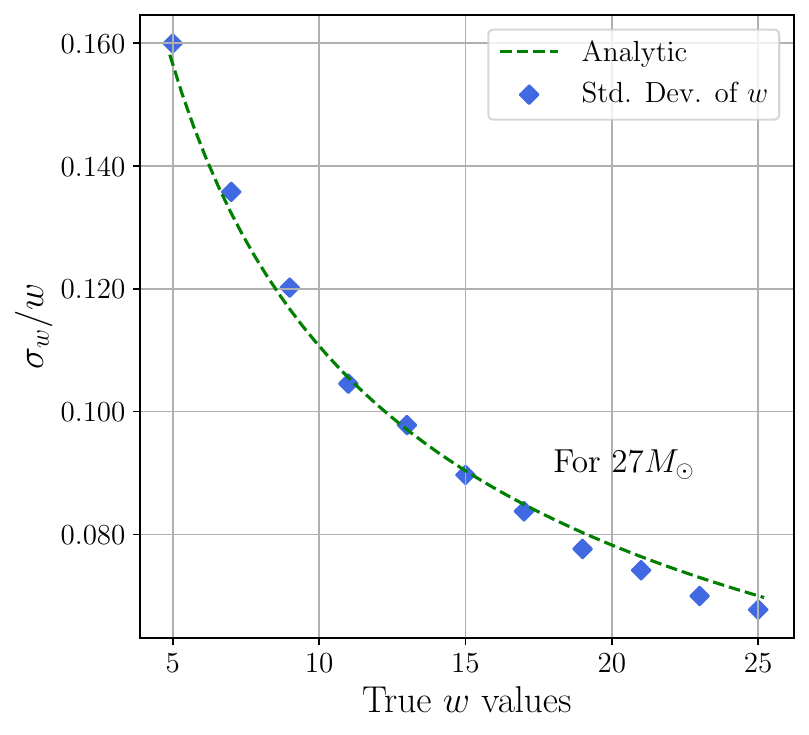}
    \includegraphics[width=0.24\linewidth]{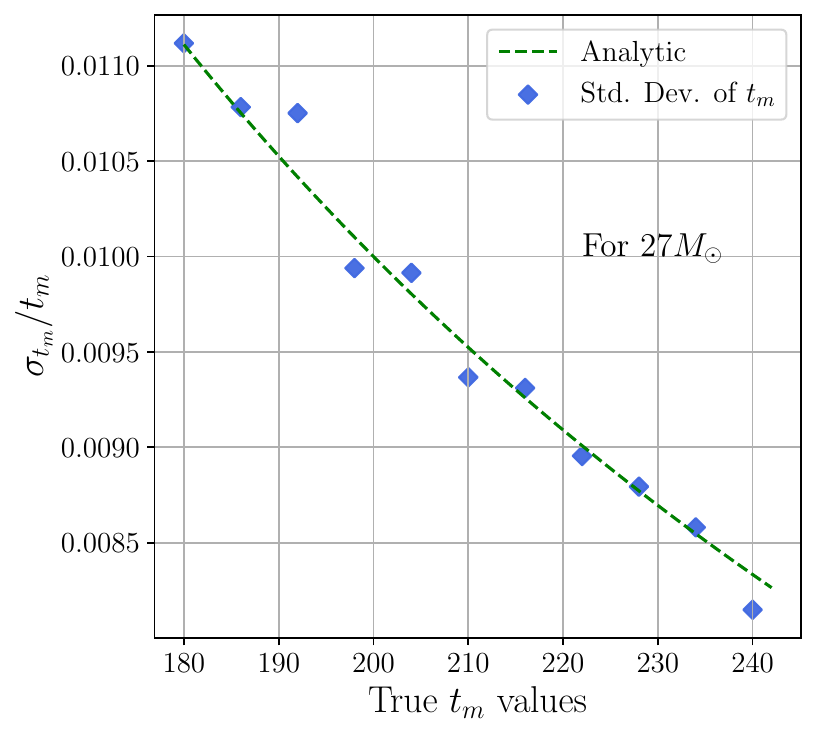}
    \includegraphics[width=0.24\linewidth]{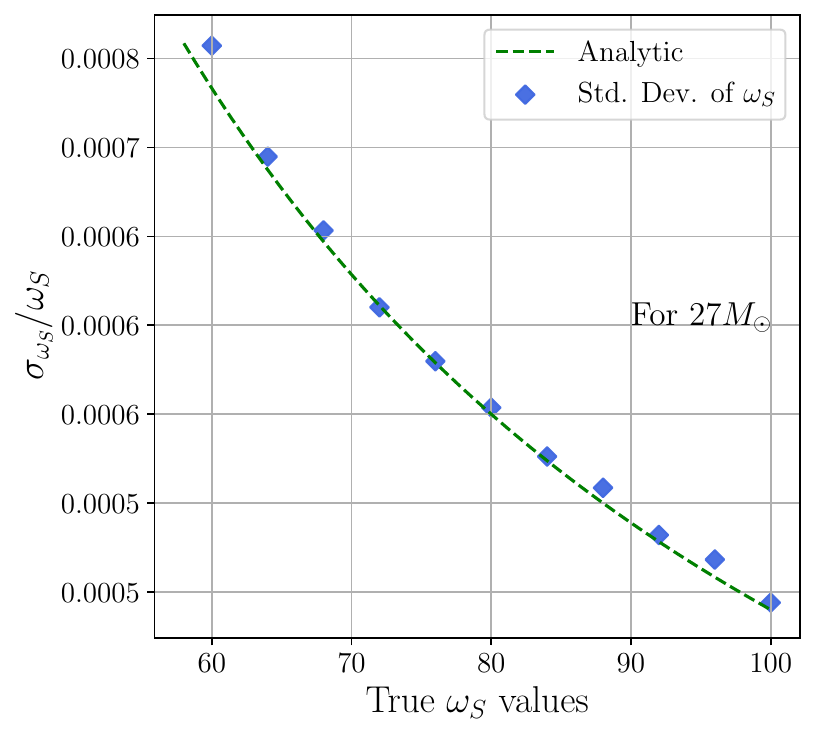}
    
    \caption{Fractional error in the extracted SASI parameters as a function of the initial parameter value, for 20$\rm{M}_{\odot}$ (top row) and 27$\rm{M}_{\odot}$ (bottom row). Panels from left to right show the amplitude $\mA$, width $w$, max-time $t_m$, and SASI frequency $\os$. Each point represents the standard deviation obtained from synthetic data of the {\sc IceCube} event rate; the dashed lines are power-law fits. The amplitude, max-time, and frequency follow inverse scaling of the parameters, while the width exhibits $\sigma_w/w \propto 1/\sqrt{w}$.}
    \label{fig:Std_dev_SASI_params}
\end{figure*}

The resulting joint and marginalized posterior distributions of the extracted amplitude $\mA$ and width $w$ for the 20$\rm{M}_{\odot}$ and 27$\rm{M}_{\odot}$ cases are shown in Fig.\,\ref{fig:Dist_of_extracted_amp_width}. The contours denote the 68\% and 95\% confidence regions, and the \emph{true} parameter values are shown by a star. The posteriors remain centered around the true values, showing that the SASI parameters can be recovered after including statistical and background noise relevant for {\sc IceCube}. To quantify the uncertainty in an extracted parameter, we compute the standard deviation $\sigma$ from these distributions and evaluate the fractional error for a SASI parameter over different initial values, while keeping the other parameters fixed, as shown in Fig.\,\ref{fig:Std_dev_SASI_params}. The dashed lines in the figure are power-law fits, which determine the fall-off index and normalization. The fluctuations in the curves around the predicted power-law behavior are due to the statistical nature of the generated realizations. For a Galactic SN at a benchmark distance of 10\,kpc, we find that the most accurately recovered parameters are the SASI frequency at sub-percent precision and the max-time at the percent level. In contrast, the amplitude and width are recovered with fractional errors at the level of a few to a few tens of percent. For the amplitude, we find $\sigma_{\mA}/\mA \propto 1/\mA$.  A similar behavior is seen for the max-time $t_m$ and the SASI frequency $\os$. For the width $w$, the fractional error follows approximately $\sigma_w/w \propto 1/\sqrt{w}$.  We explain this behavior by calculating the Fisher information matrix ${\sf F}$ for all the parameters, where its elements are related to the covariance of the parameters as ${\rm Cov}\propto ({\sf F}_{mn})^{-1}$ as derived in the Appendix.

\section{Summary and Outlook}
\label{sec:discussion_outlook}

In this work, we have developed an effective four-parameter ansatz for the SASI-modulation in supernova neutrinos, characterized by the amplitude $\mA$, max-time $t_m$, width $w$, and frequency $\os$. We validated the ansatz using the 20$\rm{M}_{\odot}$ and 27$\rm{M}_{\odot}$ simulations of the Garching group, across different viewing directions. The reconstructed waveforms show that the ansatz captures the prominent SASI-modulation very well, especially near the peak of SASI activity, with deviations visible near the onset and termination. Finally, we show that for synthetically generated data, including statistical and background noise relevant for {\sc IceCube}, the initial SASI parameters from a Galactic SN can be recovered with negligible bias and acceptable uncertainty.

The fractional-error estimates show that the SASI frequency is inferred very precisely, with sub-percent accuracy for $\os\simeq79$ Hz. The max-time $t_m$ is also well-determined, typically at the percent level, while SASI amplitude $\mA$ and SASI duration $w$ are less tightly constrained, with uncertainties at the $\sim1-10\%$ level. The upshot is that the location of the SASI peak and the SASI frequency are expected to be robustly measurable, whereas the amplitude and duration exhibit larger uncertainty.

With next-generation detectors like {\sc IceCube}-Gen2, the SASI detection prospects will improve. With an effective volume of about 7.9\;km$^3$, nearly eight times the volume of {\sc IceCube}, its in-ice component is expected to record an order of magnitude more neutrino interactions and detect sources about five times fainter. The increased number of detected counts per bin will improve sensitivity to supernovae at larger Galactic distances~\cite{Beise:2023naa}. Most relevant for our analysis, the improved time resolution of 10\;ns will allow finer features of the SASI-modulation to be resolved. With {\sc IceCube}-Gen2, we expect new avenues to observe a future Galactic supernova.

\begin{acknowledgments}
We are grateful to Hans-Thomas Janka, Daniel Kresse, and the Garching simulation group for granting us access to their simulation data. This work is supported by the Dept.~of Atomic~Energy~(Govt.~of~India) research project RTI 4012. DM acknowledges Kapil Ghadiali and Rupak Majumdar for their help with using the high-performance computing facility at the Department of Theoretical Physics, TIFR. MR acknowledges the support from Dept.~of Atomic~Energy~(Govt.~of~India).
\end{acknowledgments}

\appendix

\section{Fisher Information for SASI}

To study the dependence of the fractional error on the parameters, we write the event rate in {\sc IceCube} as
\begin{equation}
  R(t;\boldsymbol{\theta})=
  R_\nu(t)\,\Bigl[1 + \mA\,G(t)\,\sin(\os t)\Bigr]
  + R_{\rm bkg},
  \label{eq:SASI_rate_model}
\end{equation}
where $G(t)$ is a Gaussian envelope with $w$ and $t_m$ as parameters, defined in Eq.\,\ref{eq:fitting_function_eqn}. We collect the parameters into $\boldsymbol{\theta}=(\mA,\,w,\,t_m,\,\os).$
In the time-binned analysis, we consider bins centered at $t_i$ with width $\Delta t$ and define the expected number of counts in bin $i$ as
\begin{equation}
  N^{th}_i(\boldsymbol{\theta})= R(t_i;\boldsymbol{\theta})\,\Delta t.
\end{equation}
\subsection{Gaussian Likelihood and Fisher Information}

We assume that the observed counts in each bin can be approximated as
Gaussian-distributed around the model prediction,
\begin{equation}
  N_i \sim \mathcal{N}\!\bigl(N^{th}_i(\boldsymbol{\theta}),\,\sigma_i^2\bigr),
\end{equation}
with known variances $\sigma_i^2$.  The Gaussian log-likelihood
is then
\begin{equation}
  \ln\mathcal{L}(\boldsymbol{\theta})=
  -\frac{1}{2}\sum_i
  \left[
    \frac{\bigl(N_i-N^{th}_i(\boldsymbol{\theta})\bigr)^2}{\sigma_i^2}
    + \ln(2\pi\sigma_i^2)
  \right].
\end{equation}
The Fisher information matrix is defined by~\cite{Narsky:2014fya},
\begin{equation}
  {\sf F}_{mn}\equiv-\Bigl\langle
  \frac{\partial^2 \ln\mathcal{L}}{\partial\theta_m\,\partial\theta_n}
  \Bigr\rangle_{\text{data}|\boldsymbol{\theta}}.
\end{equation}
For the Gaussian likelihood with $\sigma_i$ independent of
$\boldsymbol{\theta}$, this reduces to
\begin{align}
      {\sf F}_{mn}& =\sum_i  \frac{1}{\sigma_i^2}\,
  \frac{\partial N^{th}_i}{\partial\theta_m}\,
  \frac{\partial N^{th}_i}{\partial\theta_n} \nonumber \\ 
  &=\Delta t^2\sum_i\frac{1}{\sigma_i^2}\,
  \frac{\partial R(t_i;\boldsymbol{\theta})}{\partial\theta_m}\,
  \frac{\partial R(t_i;\boldsymbol{\theta})}{\partial\theta_n}.
  \label{eq:Fisher_Gaussian_discrete}
\end{align}
The inverse of the Fisher matrix sets the Cram\'er-Rao lower bound on the parameter covariance as
\begin{equation}
  \mathrm{Cov}(\boldsymbol{\theta}) \geq {\sf F}^{-1}\implies
  \sigma_{\theta_m}^2 \simeq ({\sf F}^{-1})_{mm}.
  \label{eq:Cramer_rao_bound}
\end{equation}
We use the scaling of the Fisher elements to infer the scaling behavior of the parameter variances $\sigma_{\theta_m}$.

\subsection{Scaling of Diagonal Fisher Elements}

To derive the approximate scaling of the parameter variances, we focus on the diagonal Fisher elements ${\sf F}_{mm}$ and make certain assumptions as follows:
\begin{enumerate}
  \item The variance in each bin is set by the smooth signal plus background and varies slowly over the SASI envelope, $\sigma_i^2 \simeq R(t_i;\boldsymbol{\theta})\,\Delta t$. Furthermore, $R_\nu(t)$ does not vary during the SASI period, and we treat it as a constant $R_0$. 
        
  \item The envelope $G(t)$ is localized around $t\,\simeq\,t_m$ with width $\sim w$, so the sums over $i$ can be approximated by integrals over a Gaussian centered at $t_m$ in the limit of a large number of time bins.

\end{enumerate}
Under these assumptions, the overall time-dependence in each ${\sf F}_{mm}$ can be expressed as Gaussian integrals that determine how the Fisher element scales with $\mA$, $w$, $t_m$, and $\os$.
Substituting $\sigma_i^2 = R_i\,\Delta t$ in Eq.\,\ref{eq:Fisher_Gaussian_discrete} and going to the continuum limit $\sum_i \Delta t \to \int dt$, the diagonal elements can be schematically written as
\begin{equation}
  {\sf F}_{mm}=\int dt\;\dfrac{1}{R(t;\boldsymbol{\theta})}
  \Bigg(\frac{\partial R(t;\boldsymbol{\theta})}{\partial\theta_m}\Bigg)^2,
\end{equation}
keeping only the leading parameter dependencies. It is convenient to define variables 
\begin{equation}
    a\equiv \mA G(t)\sin(\os t),\;
  x \equiv\frac{\sqrt{\pi}\mA^2 R_0}{4w},\;
  y \equiv e^{-(\os w)^2}.
\end{equation}
These combinations arise from the Gaussian integrals that follow in the next part, where we derive the approximate scalings.

\begin{enumerate}
\item Amplitude $\mA$:
\begin{align}
  \frac{\partial R}{\partial \mA}=R_0\,G(t)\,\sin(\os t).
  \label{eq:dR_dA}
\end{align}
The derivative has no explicit $\mA$ dependence; the leading dependence enters through
\begin{equation}
  {\sf F}_{\mA\mA}=
  \int dt\,\frac{R_0G^2(t)\,\sin^2(\os t)}
  {1+a}.
\end{equation}
We are in the regime where $\mA<1$ and $|\sin|\leq1$, so we expand $1/(1+a)\approx 1-a+a^2+\dots$. Keeping the leading term gives,
\begin{align}
  {\sf F}_{\mA\mA} =\frac{\sqrt{\pi}}{2}R_0 w\Big[1-\!y\cos(2t_m\os)\Big].
\end{align}
Thus, from Eq.\,\ref{eq:Cramer_rao_bound} we obtain
\begin{align}
     \sigma_{\mA}\simeq \frac{1}{\sqrt{{\sf F}_{\mA\mA}}}.
\end{align}
For representative values of $w=25$\;ms and $\os=80$\;Hz, $y\sim e^{-4}$, so the next correction is small. Hence
\begin{equation}
  \sigma_{\mA}\approx
  \left(\frac{\sqrt{\pi}}{2}R_0 w\right)^{-1/2}\,,
 {\rm~thus~}
  \frac{\sigma_{\mA}}{\mA}\propto \frac{1}{\mA}.
\end{equation}
Higher-order terms give only subleading corrections to this scaling.

\item Width $w$:
\begin{align}
    \frac{\partial R}{\partial w}=
  R_0\,\mA\,G(t)\,\sin(\os t)\,\frac{(t - t_m)^2}{w^3}.
  \label{eq:dR_dw}    
\end{align}
The corresponding Fisher element is
\begin{equation}
  {\sf F}_{ww}=\int dt\;\frac{R_0a^2{(t-t_m)^4}}{{w^6}(1+a)}.
\end{equation}
Keeping the leading term in the denominator expansion and evaluating the Gaussian integral using {\tt Mathematica} gives,
\begin{align}
&{\sf F}_{ww}= \frac{x}{4}\bigg[6- 2y\,
\Big(4 w^4 \os^4 \nonumber\\
&\hspace{1 cm} - 12 w^2 \os^2 + 3\Big)\cos(2t_m \os)\bigg].
\end{align}
For $y\ll1$, the second term can be neglected. Thus, ${\sf F}_{ww}\approx 3x/2$ and,
\begin{align}
    \sigma_w \simeq \frac{1}{\sqrt{{\sf F}_{ww}}}\approx
  \sqrt{\frac{2}{3x}}\sim\frac{\sqrt{w}}{\mA}.
\end{align}
Therefore, the fractional error scales as
\begin{equation}
  \frac{\sigma_w}{w}\propto\frac{1}{\mA\sqrt{w}}.
\end{equation}
Even at the next order in $\mA$ the overall dependence on $w$ remains the same. At fixed $\mA$, this gives $\sigma_w/w \propto 1/\sqrt{w}$.

\item Max-time $t_m$: 
\begin{align}
  \frac{\partial R}{\partial t_m}= R_0\,\mA\,G(t)\,\sin(\os t)\,\frac{(t - t_m)}{w^2},
  \label{eq:dR_dtm}
\end{align}
leading to the Fisher element,
\begin{align}
  {\sf F}_{t_m t_m}=
\int dt\; \frac{R_0a^2 {(t - t_m)^2}}{w^4(1+a)} .
\end{align}
At leading order, we get
\begin{align}
    \phantom{xxxx}{\sf F}_{t_m t_m}\!=x\Big[1\!+y(2w^2\os^2\!-\!1)\cos(2t_m\os)\Big].
\end{align}
Neglecting the $y$ term, there is no explicit dependence on $t_m$ itself (when $t_m$ is well within the SASI active period). Therefore, ${\sf F}_{t_m t_m}\approx x$ and,
\begin{equation}
   \sigma_{t_m}\simeq\frac{1}{\sqrt{{\sf F}_{t_m t_m}}}
  \approx
  \frac{1}{\sqrt{x}}\sim \frac{\sqrt{w}}{\mA}.
\end{equation}
This implies that the absolute uncertainty on $t_m$ is approximately constant as $t_m$ is varied, so the fractional error (for fixed $\mA$ and $w$) scales as
\begin{equation}
  \frac{\sigma_{t_m}}{t_m}\propto\frac{1}{t_m}.
\end{equation}

\item SASI frequency $\os$:
\begin{equation}
  \frac{\partial R}{\partial \os}  =
  R_0\,\mA\,G(t)\,t\,\cos(\os t).
  \label{eq:dR_domega}
\end{equation}
Thus, the Fisher element is
\begin{equation}
  {\sf F}_{\os\os}= 
  \int dt\,\frac{R_0t^2(\mA^2G^2(t)-a^2)}{(1+a)}. 
\end{equation}
Keeping the leading term in $\mA$, the integral becomes,  
\begin{align}
&{\sf F}_{\os\os}=xw^2\bigg[(2 t_m^2 + w^2)\nonumber\\
&\quad+ y\Big((2 t_m^2 + w^2 - 2 \os^2 w^4)\cos(2t_m\os)\nonumber\\
&\quad- 4 \os t_m w^2 \sin(2t_m\os)
\Big)\bigg].
\end{align}
For fixed $\mA$, $t_m$, and $w$, the sine and cosine terms are suppressed by $y$. Hence, ${\sf F}_{\os\os}\approx x\,w^2(2t_m^2+w^2)$ and,
\begin{equation}
    \sigma_{\os}\approx
  \frac{1}{w\sqrt{x(2t_m^2+w^2)}}.
\end{equation}
Thus, the fractional error scales as
\begin{equation}
  \frac{\sigma_{\os}}{\os}\propto\frac{1}{\os}.
\end{equation} 

\end{enumerate}

Collecting these results, we recover the trends observed in Fig.\,\ref{fig:Std_dev_SASI_params}. We note that these scalings are derived under simplifying assumptions and should be interpreted as approximate, since the fitted parameters obtained in our procedure do not always lie strictly within the range in which these assumptions hold.

\linespread{0.93}
\bibliographystyle{JHEP}
\bibliography{References}
\end{document}